\pdfoutput=1 

\documentclass[preprint,superscriptaddress,nofootinbib,longbibliography]{revtex4-1}
\usepackage[T1]{fontenc}
\usepackage[latin9]{inputenc}
\usepackage{geometry}
\usepackage{amsmath}
\usepackage{amssymb}
\usepackage{graphicx}
\usepackage{babel}
\usepackage{float}
\usepackage{xcolor}

\begin{document}
\title{Cosmic Evolution of Lepton Flavor Charges}
\newcommand{\affUFABC}{Centro de Ci\^encias Naturais e Humanas\;\;\\
	Universidade Federal do ABC, 09.210-170,
	Santo Andr\'e, SP, Brazil}

\author{Chee Sheng Fong}
\email{sheng.fong@ufabc.edu.br}
\affiliation{\affUFABC}

\begin{abstract}
In the early Universe above the weak scale, both baryon $B$ and lepton $L$ numbers are violated by nonperturabive effects in the Standard Model while $B-L$ remains conserved. Introducing new physics which violates perturbatively $L$ and/or $B$, one can generate dynamically a nonzero $B-L$ charge and hence a nonzero $B$ charge. In this work,
we focus on the former scenario which is also known as leptogenesis.
We show how to describe the evolutions of lepton flavor charges taking into account the complete Standard Model lepton flavor and spectator effects in a unified and lepton flavor basis-independent way. The recipe we develop can be applied to any leptogenesis model with arbitrary number of new scalars carrying nonzero hypercharges and is valid for cosmic temperature ranging from $10^{15}$ GeV down to the weak scale.
We demonstrate that in order to describe the physics in a basis-independent manner and to include lepton flavor effect consistently it is necessary to describe both left-handed and right-handed lepton charges in terms of density matrices. This is a crucial point since physics should be basis independent.  
As examples, we apply the formalism to type-I and type-II leptogenesis models where in the latter case, a flavor-covariant formalism is indispensable. 
\end{abstract}

\maketitle
\flushbottom

\section{Introduction}

In the early Universe, if the cosmic temperature is above the weak scale, the thermal bath contains all the degrees of freedom of the Standard Model (SM) and perhaps other new physics degrees of freedom
as well if they are kinematically accessible. To generate a cosmic baryon asymmetry dynamically (baryogenesis), one needs to violate at least the baryon
number $B$ of the SM. Above the weak scale when the SM $B$-violating
process is in thermal equilibrium \cite{Kuzmin:1985mm}, one needs
to identify other charges which are not in thermal equilibrium such that the charge is effectively conserved and can remain nonzero. In the SM, one identifies the baryon minus lepton number $B-L$ as the exactly conserved charge. If one introduces new physics which perturbatively violates $B-L$, together with violation of charge $C$ and charge parity $CP$, a nonzero $B-L$ charge can be dynamically generated. Since $B$ is not orthogonal to $B-L$, we have
%%%
\begin{eqnarray}
B & = & c\left(B-L\right),\label{eq:B-L_to_B_general}
\end{eqnarray}
%%%
with $c\neq0$, implying a nonzero $B$ is generated as well.
After baryogenesis is completed, i.e., $(B-L)$-violating interaction goes
out of equilibrium, while $B-L$ charge remains conserved, it is important to note that, since $B$ is not a conserved charge, it can (and in general will) evolve with cosmic temperature. In other words, the
coefficient $c$ that relates $B$ and $B-L$ in eq. \eqref{eq:B-L_to_B_general} is temperature dependent
since it depends on the effective charges of the thermal bath. How can effective charges arise in the early Universe? They arise as the cosmic temperature increases when some of the SM interactions go out of equilibrium. In principle, baryogenesis does not have to go through $B-L$ but can proceed through other effective charges ${\cal Q}$ which are not completely orthogonal
to $B$ \cite{Fong:2015vna}
%%%
\begin{eqnarray}
B & = & \sum_{{\cal {\cal Q}}}c_{{\cal Q}}{\cal Q},
\end{eqnarray}
%%%
with $c_{{\cal Q}}\neq0$. In ref. \cite{Fong:2020fwk}, we have classified all effective charges of the SM and its minimal supersymmetric extension, 16 in the former and 18 in the latter and this opens up a new avenue
for baryogenesis.

In this work, we focus on baryogenesis scenario through the violation
of $B-L$ which can come from perturbative interaction which violates $L$ and/or $B$. We consider the former scenario, which
is also known as leptogenesis \cite{Fukugita:1986hr}. First of all,
we show that in order describe leptogenesis in a \emph{basis-independent}
manner one needs to describe both the number asymmetries in lepton
doublet $\ell$ and singlet $E$ in term of matrices of number densities in their respective
flavor spaces (we will denote them densities matrices) \cite{Sigl:1992fn}.
It is of fundamental importance since physics should not depend on a particular basis. While the
computation of leptogenesis is usually carried out in a charged lepton
mass basis, one should be cautious that this description has limited
validity, and in particular, if the result is basis dependent, then it is a red flag that something must be wrong. In this flavor-covariant formalism \cite{Beneke:2010dz,Garbrecht:2013bia}, the SM
lepton flavor effect is consistently taken into account.\footnote{Ref. \cite{BhupalDev:2014oar} develops a flavor-covariant formalism which takes into account the flavors of left-handed SM leptons as well as the massive right-handed neutrinos in the type-I seesaw model.} 
With the effective charges identified in ref. \cite{Fong:2020fwk}, we are able to include the complete spectator effects due to quark Yukawa and SM sphaleron interactions in a unified manner, which to our knowledge has not been carried out before. (See ref. \cite{Garbrecht:2014kda}, in which the spectator effects related to tau and bottom-quark Yukawa interactions are investigated.)
In ref. \cite{Blanchet:2011xq}, asymmetry in $E$ is not taken in account, and as a result, one cannot obtain a fully basis-independent result. In ref. \cite{Beneke:2010dz,Garbrecht:2013bia}, asymmetry in singlet $E$ is considered while other spectator effects \cite{Buchmuller:2001sr,Nardi:2005hs}
pertaining to quark Yukawa and SM sphaleron interactions are not considered.

This article is organized as follows. In Section \ref{sec:effective_charges},
we review the effective symmetries and charges of the SM in the early Universe. In Section
\ref{sec:Lepton-flavor-effect}, we write down the flavor-covariant Boltzmann equations, taking into account the complete lepton flavor and spectator
effects due to quark Yukawa and the SM sphaleron interactions. These results are completely general, and, together with the equations in Appendix
\ref{app:quark_relations}, can be applied to any leptogenesis model (with arbitrary number of new scalars carrying nonzero hypercharges)
for cosmic temperature ranging from $10^{15}$ GeV down to the weak scale.
In Section \ref{sec:Applications}, we apply our results to type-I and type-II leptogenesis models. Finally, we conclude in Section \ref{sec:Conclusions}.
In Appendix \ref{app:matrix_number_densities}, we discuss how number density asymmetry matrices are related to matrices of chemical potentials; in Appendix \ref{app:flavor_kinetic_equations}, we show how the flavor-covariant structure can be derived using Sigl-Raffelt formalism \cite{Sigl:1992fn}; and in
Appendix \ref{app:trans_temps}, we discuss how to determine the transition temperatures related to spectator effects.

\section{Effective symmetries and charges\label{sec:effective_charges}}

In the early Universe, due to the additional scale related to cosmic expansion, one should consider effective symmetries and charges. To illustrate this point, let us consider the early Universe which is
dominated by radiation energy density $\rho_{r}\propto T^{4}$ with temperature $T$ and is expanding with the
Hubble rate ${\cal H}\propto\sqrt{\rho_{r}}/M_{{\rm Pl}}\propto T^{2}/M_{{\rm pl}}$,
where $M_{{\rm Pl}}=1.22\times10^{19}$ GeV is the Planck scale. Taking
all particles to be massless, the interaction rates among the particles
have to scale as $T$. At sufficiently high $T$, all
of those interactions will be slower than the Hubble rate. In this
case, if one assigns a quantum number or charge to each type of particle,
the charge will be effectively conserved since all particle-number-changing
processes are out of thermal equilibrium (effectively do not occur within a Hubble time). In the SM, with three
families $\alpha=1,2,3$ of quark $Q_{\alpha}$ and lepton $\ell_{\alpha}$
doublets, charged lepton $E_{\alpha}$, up-type $U_{\alpha}$ and
down-type $D_{\alpha}$ quarks singlets, and a Higgs doublet, one will
expect to have up to 16 effective charges or the associated global
$U(1)$ symmetries. 
One can conveniently choose linear combinations of $U(1)$ charges which are subsequently broken as the cosmic temperature decreases. This choice leads to $U(1)_x$ with \cite{Fong:2020fwk}\footnote{In the minimal supersymmetric SM, there are 18 effective symmetries and corresponding charges.}
%%%
\begin{eqnarray}
x & = & \left\{ t,u,B,\text{\ensuremath{\tau}},u-b,u-c,\mu,B_{3}-B_{2},u-s,B_{3}+B_{2}-2B_{1},u-d,e,B/3-L_{\alpha},Y\right\} ,\label{eq:SM_global_symmetries}
\end{eqnarray}
%%%
where we have denoted the charge associated to each type of particle
as $\left\{ U_{1},U_{2},U_{3}\right\} =\left\{ u,c,t\right\} $, $\left\{ D_{1},D_{2},D_{3}\right\} =\left\{ d,s,b\right\} $, and
$\left\{ E_{1},E_{2},E_{3}\right\} =\left\{ e,\mu,\tau\right\} $
and $B_{\alpha}$ refers to baryon flavor number with the total baryon
number $B=B_{1}+B_{2}+B_{3}$, while $L_{\alpha}$ refers to lepton
flavor number with the total lepton number $L=L_{1}+L_{2}+L_{3}$.
Out of these 16 $U(1)_{x}$, only the last four remain exact before
the electroweak (EW) symmetry breaking: hypercharge gauge symmetry
$Y$ and the three $B/3-L_{a}$ accidental (global) symmetries. The
rest of the effective symmetries are broken by the Yukawa and nonperturbative sphaleron interactions. 

In the absence of neutrino mass, the SM Lagrangian contains four accidental
$U(1)$ symmetries: the total baryon number $U(1)_{B}$ and three
lepton flavors $U(1)_{L_{\alpha}}$.
There are fewer actual accidental symmetries of the SM due to the Adler-Bell-Jackiw anomaly.
We can determine if any of the accidental symmetry $U(1)_{x}$ is
preserved from its anomaly coefficient associated with the triangle
diagram $U(1)_{x}-SU(N)-SU(N)$
%%%
\begin{eqnarray}
A_{xNN} & \equiv & \sum_{i}c(R_{i})g_{i}q_{i}^{x},
\end{eqnarray}
%%%
where the sum is over all fermions $i$ of degeneracy $g_{i}$, charge
$q_{i}^{x}$ under $U(1)_{x}$, and representation $R_{i}$ under $SU(N\geq2)$
gauge group with $c(R_{i})=\frac{1}{2}$ in the fundamental representation
and $c(R_{i})=N$ in the adjoint representation. Since the contribution
of each fermion $i$ to the $SU(N)$ sphaleron-induced effective operator
is proportional to $c(R_{i})$, the effective operator is given by \cite{Fong:2020fwk}
%%%
\begin{eqnarray}
{\cal O}_{{\rm SU(N)}} & \sim & \prod_{i}\Psi_{i}^{2g_{i}c(R_{i})}.\label{eq:sphaleron_operator}
\end{eqnarray}
%%%
In the SM, we see that $U(1)_{B}$ and $U(1)_{L_{\alpha}}$ are anomalous
\cite{tHooft:1976rip} with anomaly coefficients
%%%
\begin{eqnarray}
A_{B22} & \equiv & \frac{1}{2}\times3\left(3\times\frac{1}{3}\right)=\frac{3}{2},\\
A_{L_{\alpha}22} & \equiv & \frac{1}{2}\times1=\frac{1}{2}.
\end{eqnarray}
%%%
Out of four anomalous symmetries, one can form three linear combinations
which are anomaly free. It is convenient to choose the following three
anomaly-free symmetries $U(1)_{B/3-L_{\alpha}}$ we mentioned
earlier. Then, the anomalous symmetry $U(1)_{B+L}$ with anomaly coefficient
%%%
\begin{eqnarray}
A_{(B+L)22} & = & \frac{1}{2}\times3\left(3\times\frac{1}{3}+1\right)=3.
\end{eqnarray}
%%%
From eq. (\ref{eq:sphaleron_operator}), one obtains the EW sphaleron effective operator
%%%
\begin{eqnarray}
{\cal O}_{{\rm SU(2)}} & \sim & \prod_{\alpha=1}^{3}Q_{\alpha}Q_{\alpha}Q_{\alpha}\ell_{\alpha}.
\end{eqnarray}
%%%
The operator above violates only $U(1)_{B}$, and the interaction due
to this operator is in thermal equilibrium \cite{Kuzmin:1985mm} from
$T_{B}\sim2\times10^{12}\,{\rm GeV}$ \cite{Garbrecht:2014kda} up
to $T_{B-}\sim132$ GeV \cite{DOnofrio:2014rug}.

The SM quark Yukawa terms are given by
%%%
\begin{eqnarray}
-{\cal L} & \supset & \left(y_{U}\right)_{ab}\overline{U_{a}} Q_{b}\epsilon H+\left(y_{D}\right)_{ab}\overline{D_{a}} Q_{b}H^{*}+{\rm H.c.},
\end{eqnarray}
%%%
where the $SU(2)_{L}$ contraction between the left-handed quark $Q_{b}$ and
the Higgs $H$ doublets is shown explicitly with the $SU(2)$ antisymmetric
tensor $\epsilon_{01}=-\epsilon_{10}=1$. If these terms are absent,
one has a chiral symmetry $U(1)_{\chi}$ where $q_{Q_{a}}^{\chi}=-q_{U_{a}}^{\chi}=-q_{D_{a}}^{\chi}\equiv q$.
Nevertheless, this chiral symmetry is anomalous with
%%%
\begin{eqnarray}
A_{\chi33} & = & \frac{1}{2}\times3\left(2\times q+q+q\right)=6q.
\end{eqnarray}
%%%
From eq. (\ref{eq:sphaleron_operator}), one can construct the QCD sphaleron effective operator as \cite{Moore:1997im}
%%%
\begin{eqnarray}
{\cal O}_{{\rm SU(3)}} & \sim & \prod_{\alpha=1}^{3}Q_{\alpha}Q_{\alpha}U_{\alpha}^{c}D_{\alpha}^{c}.
\end{eqnarray}
%%%
The operator above violates the chiral symmetry $U(1)_{u}$, and the interaction due
to this operator is in thermal equilibrium for $T\lesssim T_{u}\sim2\times10^{13}$
GeV \cite{Garbrecht:2014kda}.

The rest of the effective symmetries in eq. \eqref{eq:SM_global_symmetries}
are broken when the corresponding Yukawa interactions get into thermal
equilibrium, starting from the one involving top Yukawa, tau Yukawa and so on. We can estimate the temperature $T_{x}$ in which $U(1)_{x}$
is broken from the condition when the $U(1)_{x}$-violating rate is
equal to the Hubble rate $\Gamma_{x}\left(T_{x}\right)={\cal H}\left(T_{x}\right)$
and obtain \cite{Fong:2020fwk}
%%%
\begin{eqnarray}
T_{t} & \sim & 1\times10^{15}\,{\rm GeV},\nonumber \\
T_{u} & \sim & 2\times10^{13}\,{\rm GeV},\nonumber \\
T_{B} & \sim & 2\times10^{12}\,{\rm GeV},\nonumber \\
T_{\tau} & \sim & 4\times10^{11}\,{\rm GeV},\nonumber \\
T_{u-b} & \sim & 3\times10^{11}\,{\rm GeV},\nonumber \\
T_{u-c} & \sim & 2\times10^{10}\,{\rm GeV},\nonumber \\
T_{\mu} & \sim & 10^{9}\,{\rm GeV},\label{eq:transition_temperatures}\\
T_{B_{3}-B_{2}} & \sim & 9\times10^{8}\,{\rm GeV},\nonumber \\
T_{u-s} & \sim & 3\times10^{8}\,{\rm GeV},\nonumber \\
T_{B_{3}+B_{2}-2B_{1}} & \sim & 10^{7}\,{\rm GeV},\nonumber \\
T_{u-d} & \sim & 2\times10^{6}\,{\rm GeV},\nonumber \\
T_{e} & \sim & 3\times10^{4}\,{\rm GeV},\nonumber 
\end{eqnarray}
%%%
and we have assumed thermalization at $T \sim 10^{15}$ GeV \cite{Davidson:2000er,Harigaya:2013vwa}.
In principle, one will need to track the evolutions of all the effective
charges, starting from some initial condition. For instance, after
reheating at the end of inflation with temperature $T_{{\rm RH}}$,
we can take the initial condition to be when all the effective charges
are zero. The charge density associated to each effective charge can
be written as
%%%
\begin{eqnarray}
n_{\Delta x} & = & \sum_{i}q_{i}^{x}n_{\Delta i},\label{eq:charge_density}
\end{eqnarray}
%%%
where the number density asymmetry of particle $i$ is defined as
$n_{\Delta i}\equiv n_{i}-n_{\bar{i}}$ where $n_{i}\,(n_{\bar{i}})$
is the number density of particle $i$ (antiparticle $\bar{i}$).
In this case, the initial condition will be $n_{\Delta x}\left(T_{{\rm RH}}\right)=0$
for all the charges. One should then track the evolutions of all the
$n_{\Delta x}\left(T\right)$ with the Boltzmann equations including
all the SM interactions. To generate some nonzero charges,
the three Sakharov conditions should be fulfilled \cite{Sakharov:1967dj}:
\begin{itemize}
\item violation of $U(1)_{x}$,
\item $C$ and $CP$ violation corresponding to the process violating $U(1)_{x}$,
\item out-of-equilibrium condition for the process violating $U(1)_{x}$.
\end{itemize}
If the Sakharov conditions are not met for any of the charges in eq. \eqref{eq:SM_global_symmetries},
one will always have $n_{\Delta x}=0$. If the Sakharov conditions
are met for some of the $U(1)_{x}$ (this does not happen in the SM and hence
physics beyond the SM is required), one will have $n_{\Delta x}\left(T_{g}\right)\neq0$,
where $T_{g}$ is the temperature when the charge $n_{\Delta x}$
is being generated. If all other $U(1)_{y\neq x}$ remain effective,
we have $n_{\Delta y}=0$, while for $U(1)_{y\neq x}$ which are not
effective, we will necessarily have $n_{\Delta y}\propto n_{\Delta x}$.
This does not necessarily imply that $n_{\Delta y}\neq0$ since the
constant of proportionality can be zero; i.e., $U(1)_{y}$ and $U(1)_{x}$
are orthogonal to each other. At $T_{B-}<T<T_{e}$, since $U(1)_{B}$
is not effective, we can construct the baryon charge density from
eq. (\ref{eq:charge_density}) as
%%%
\begin{eqnarray}
n_{\Delta B} & = & c\sum_{\alpha}n_{\Delta(B/3-L_{\alpha})},
\end{eqnarray}
%%%
where we have assumed zero hypercharge density $n_{\Delta Y}=0$.
The coefficient $c$ is not zero since $B$ and $B/3-L_{\alpha}$
are not orthogonal to each other. With the SM degrees of freedom and
assuming that the EW sphaleron interaction freezes out at 132 GeV
after the EW symmetry breaking at 160 GeV \cite{DOnofrio:2014rug},
we obtain
%%%
\begin{eqnarray}
n_{\Delta B}\left(T_{B-}\right) & = & \frac{30}{97}\sum_{a}n_{\Delta(B/3-L_{a})}\left(T_{B-}\right),\label{eq:B-L_to_B}
\end{eqnarray}
%%%
where we have excluded the top-quark contribution.

Next, we will review briefly how to relate the number density asymmetries
of the SM particles to their corresponding chemical potentials. Since
all the SM particles participate in the gauge interactions, they can thermalize at a cosmic temperature $T\lesssim10^{15}$ GeV \cite{Davidson:2000er,Harigaya:2013vwa}
and follow the equilibrium phase-space distribution
%%%
\begin{eqnarray}
f_{i} & = & \frac{1}{e^{\frac{{\cal E}_{i}-\mu_{i}}{T}}+\xi_{i}},
\label{eq:phase_space}
\end{eqnarray}
%%%
where ${\cal E}_{i}$ is the energy of particle $i$, $\mu_{i}$ is its
chemical potential, and $\xi_{i}=1(-1)$ for $i$ a fermion (boson).
For gauge bosons, their numbers are not conserved, and their chemical
potentials are zero. For the rest of the SM particles, due to the scatterings with the gauge bosons, the chemical potential of an antiparticle
is related to the corresponding particle by a negative sign $\mu_{\bar{i}}=-\mu_{i}$.
To take into account flavor correlation of particle $i$, one can generalize $\mu_{i}$ to a matrix in its flavor space. (See Appendix
\ref{app:matrix_number_densities} for details.) In this work, since we are interested in the lepton flavor effect, we will generalize $\mu_{\ell}$ and $\mu_{E}$ to matrices in their lepton flavor spaces (see the next section).

Integrating the phase space distribution eq. \eqref{eq:phase_space} over 3-momentum, at leading
order in $\left|\mu_{i}\right|/T\ll1$ (assuming that the number density asymmetries of the SM particles are much smaller than their total number densities in the early Universe in accordance with observation),
the number density asymmetries are linearly proportional to their respective chemical potentials
%%%
\begin{eqnarray}
n_{\Delta i} & \equiv & \int \frac{d^3p}{(2\pi)^3} (f_i - f_{\bar i}) =
\frac{T^{2}}{6}g_{i}\zeta_{i}\mu_{i},
\end{eqnarray}
%%%
where $g_{i}$ is the gauge degrees of freedom and $\zeta_{i}=1(2)$ for $i$ a massless fermion (boson).\footnote{For a particle $i$ with mass $m_{i}$, $\zeta_{i}=\frac{6}{\pi^{2}}\int_{m_{i}/T}^{\infty}dx\,x\sqrt{x^{2}-m_{i}^{2}/T^{2}}\frac{e^{x}}{\left(e^{x}+\xi_{i}\right)^{2}}$.}
To scale out the effect of dilution purely due to the Hubble expansion,
we will normalize the matrix of number densities $Y_{i}\equiv n_{i}/s$
by the cosmic entropy density $s=\frac{2\pi^{2}}{45}g_{\star}T{{}^3}$
with $g_{\star}$ being the effective relativistic degrees of freedom of the Universe ($g_{\star}=106.75$ for the SM) and we obtain
%%%
\begin{eqnarray}
Y_{\Delta i} & \equiv & Y_{i}-Y_{\bar{i}}\equiv Y^{{\rm nor}}g_{i}\zeta_{i}\frac{2\mu_{i}}{T},\label{eq:Y_mu_i}
\end{eqnarray}
%%%
where we have defined $Y^{{\rm nor}}\equiv\frac{15}{8\pi^{2}g_{\star}}$.
Then, one can relate $Y_{\Delta i}$ to normalized charge density $Y_{\Delta x}\equiv n_{\Delta x}/s$
as \cite{Fong:2015vna}
%%%
\begin{eqnarray}
Y_{\Delta i} & = & \sum_{x}g_{i}\zeta_{i}\sum_{y}q_{i}^{y}\left(J^{-1}\right)_{yx}Y_{\Delta x},\label{eq:YDi_YDx}
\end{eqnarray}
%%%
where
%%%
\begin{eqnarray}
J_{xy} & \equiv & \sum_{i}g_{i}\zeta_{i}q_{i}^{x}q_{i}^{y}.
\end{eqnarray}
%%%
The relation above is completely general (the charges are completely fixed for any given model), and the temperature dependence appears only in $\zeta_{i}$ for particles which are not massless and in $Y_{\Delta x}\left(T\right)$,
which should be solved from the relevant Boltzmann equations. In the next section, we will discuss how to consider lepton flavor charges and their coherences with density matrices while treating the effects of baryons as spectators \cite{Buchmuller:2001sr,Nardi:2005hs}.

\section{Lepton flavor effect\label{sec:Lepton-flavor-effect}}

In the SM, we have the charged lepton Yukawa term\footnote{In the minimal supersymmetric SM, the corresponding term in the superpotential is $W\supset\left(y_{E}\right)_{\alpha\beta}E_{\alpha}^{c}\ell_{\beta}\epsilon H_{d}$.}
%%%
\begin{eqnarray}
-{\cal L} & \supset & \left(y_{E}\right)_{\alpha\beta}\overline{E_{\alpha}}\ell_{\beta}H^{*}+{\rm H.c.},
\end{eqnarray}
%%%
where $\ell_{\beta}$ and $H$ are, respectively, the left-handed lepton
and Higgs $SU(2)_{L}$ doublets while $E_{\alpha}$ is the right-handed
charged lepton $SU(2)_{L}$ singlet with family indices $\alpha,\beta=1,2,3$.
The charged lepton Yukawa coupling can be diagonalized by two unitary
matrices $U_{E}$ and $V_{E}$,
%%%
\begin{eqnarray}
\hat{y}_{E} & = & U_{E}y_{E}V_{E}^{\dagger},\label{eq:flavor_basis}
\end{eqnarray}
%%%
where $\hat{y}_{E}=\frac{1}{v}{\rm diag}\left(m_{e},m_{\mu},m_{\tau}\right)$
with $v\equiv\left\langle H\right\rangle =174$ GeV the Higgs vacuum
expectation value and $m_{e}$, $m_{\mu}$ and $m_{\tau}$ are, respectively, the electron, muon, and tau lepton masses (at certain scale). In the charged lepton mass
basis, which is also known as the (leptonic) flavor basis, we have
$E'=U_{E}E$ and $\ell'=V_{E}\ell$ where they are labeled as $\ell'=\left\{ \ell'_{e},\ell'_{\mu},\ell'_{\tau}\right\} $
and $E'=\left\{ E'_{e},E'_{\mu},E'_{\tau}\right\} $. 

In this work, since we are interested in studying the flavor coherence
of the lepton charges, from eq. (\ref{eq:Y_mu_i}), we will consider matrices of number density asymmetries of $\ell$ and $E$
(see Appendix \ref{app:matrix_number_densities} for details),
%%%
\begin{eqnarray}
Y_{\Delta\ell} & = & Y^{{\rm nor}}g_{\ell}\zeta_{\ell}\frac{2\mu_{\ell}}{T},\;\;\;\;\;Y_{\Delta E}=Y^{{\rm nor}}g_{E}\zeta_{E}\frac{2\mu_{E}}{T},
\end{eqnarray}
%%%
where $Y_{\Delta\ell}$, $Y_{\Delta E}$, $\mu_{\ell}$, and $\mu_{E}$
are $3\times3$ Hermitian matrices in the leptonic flavor spaces (one for $\ell$
and the other for $E$). 
The diagonal elements denote the number density asymmetries in the ``flavors'' for any chosen basis (not necessarily the charged lepton mass basis), while the off-diagonal elements encode the correlations between the flavors.
As we will see later, this generalization is necessary such that physics is
independent of basis. Nevertheless, a convenient basis is usually useful to interpret the physics at hand. Including the EW sphaleron \cite{Bento:2003jv} and
scatterings due to charged lepton Yukawa, the flavor-covariant Boltzmann equations can be written as \cite{Sigl:1992fn,Beneke:2010dz,Garbrecht:2013bia,Garbrecht:2014kda}\footnote{We ignore flavor oscillations which are damped by gauge interactions \cite{Beneke:2010dz}. These equations have been derived in refs. \cite{Beneke:2010dz,Garbrecht:2013bia} using the closed time path formalism. See Appendix \ref{app:flavor_kinetic_equations} for discussion on how the flavor-covariant structures can be derived from the evolution equation of a Heisenberg operator \cite{Sigl:1992fn}.}
%%%
\begin{eqnarray}
s{\cal H}z\frac{dY_{\Delta\ell}}{dz} & = & -\frac{\gamma_{{\rm EW}}}{4Y^{{\rm nor}}}\left(\frac{{\rm Tr}Y_{\Delta\ell}}{g_{\ell}\zeta_{\ell}}+3\frac{{\rm Tr}Y_{\Delta Q}}{g_{Q}\zeta_{Q}}\right)I_{3\times3}\nonumber \\
 &  & -\frac{\gamma_{E}}{2Y^{{\rm nor}}}\left\{ y_{E}^{\dagger}y_{E},\frac{Y_{\Delta\ell}}{g_{\ell}\zeta_{\ell}}\right\} +\frac{\gamma_{E}}{Y^{{\rm nor}}}y_{E}^{\dagger}y_{E}\frac{Y_{\Delta H}}{g_{H}\zeta_{H}}+\frac{\gamma_{E}}{Y^{{\rm nor}}}y_{E}^{\dagger}\frac{Y_{\Delta E}}{g_{E}\zeta_{E}}y_{E},\label{eq:BE_Yell}\\
s{\cal H}z\frac{dY_{\Delta E}}{dz} & = & -\frac{\gamma_{E}}{2Y^{{\rm nor}}}\left\{ y_{E}y_{E}^{\dagger},\frac{Y_{\Delta E}}{g_{E}\zeta_{E}}\right\} -\frac{\gamma_{E}}{Y^{{\rm nor}}}y_{E}y_{E}^{\dagger}\frac{Y_{\Delta H}}{g_{H}\zeta_{H}}+\frac{\gamma_{E}}{Y^{{\rm nor}}}y_{E}\frac{Y_{\Delta\ell}}{g_{\ell}\zeta_{\ell}}y_{E}^{\dagger},\label{eq:BE_YE}
\end{eqnarray}
%%%
where we have defined the anticommutator $\left\{ A,B\right\} \equiv AB+BA$,
$z\equiv\frac{M_{{\rm ref}}}{T}$ with an arbitrary reference mass
scale $M_{{\rm ref}}$ and ${\cal H}=1.66\sqrt{g_{\star}}T^{2}/M_{{\rm Pl}}$
is the Hubble rate for a radiation-dominated Universe. The charged
lepton Yukawa reaction density was determined in ref. \cite{Cline:1993bd,Garbrecht:2013bia}
to be $\gamma_{E}\approx5\times10^{-3}\frac{T^{4}}{6}$ 
where thermal corrections and scatterings involving gauge fields and quark fields are taken into account. The EW sphaleron reaction density was determined in ref. \cite{DOnofrio:2014rug}
to be $\gamma_{{\rm EW}}\approx18\alpha_{2}^{5}T^{4}$ where $\alpha_{2}=\frac{g_{2}^{2}}{4\pi}$
with $g_{2}$ the weak coupling. Under arbitrary flavor rotations
%%%
\begin{eqnarray}
E & \to & UE,\;\;\;\;\ell\to V\ell,\;\;\;\;\;y_{E}\to Uy_{E}V^{\dagger},\label{eq:flavor_rotations}
\end{eqnarray}
%%%
 the Boltzmann equations (\ref{eq:BE_Yell}) and (\ref{eq:BE_YE})
are manifestly covariant if 
%%%
\begin{eqnarray}
Y_{\Delta\ell} & \to & VY_{\Delta\ell}V^{\dagger},\;\;\;\;\;Y_{\Delta E}\to UY_{\Delta E}U^{\dagger}.\label{eq:transformation_leptonic_charges}
\end{eqnarray}
%%%
The above transformations can be easily ensured when constructing
the matrix of number density asymmetry as shown in Appendix \ref{app:matrix_number_densities}.
Hence, we can use the freedom above to work in \emph{any} basis while
the observables i.e. ${\rm Tr}Y_{\Delta\ell}$ and ${\rm Tr}Y_{\Delta E}$
remain unaffected by our choice of basis. For instance, we can choose $U=U_{E}$ and $V=V_{E}$
which correspond to flavor basis (\ref{eq:flavor_basis}).\footnote{Including the Renormalized Group Evolution (RGE) of charged lepton Yukawa
coupling, $U_{E}$ and $V_{E}$ will in general be scale dependent.}

In the SM, the Boltzmann equation for the evolution of total baryonic
charge $Y_{\Delta B}$ is the following \cite{Bento:2003jv}
%%%
\begin{eqnarray}
s{\cal H}z\frac{dY_{\Delta B}}{dz} & = & -\frac{3\gamma_{{\rm EW}}}{4Y^{{\rm nor}}}\left(\frac{{\rm Tr}Y_{\Delta\ell}}{g_{\ell}\zeta_{\ell}}+3\frac{{\rm Tr}Y_{\Delta Q}}{g_{Q}\zeta_{Q}}\right).\label{eq:BE_YB}
\end{eqnarray}
%%%
The additional factor of 3 comes from the fact that for each scattering,
the change of the total baryon number is $\Delta B=3$ while for the
lepton flavors we have $\Delta L_{\alpha}=1$ for each flavor. In
this work, our focus is only on the lepton flavor effect, and hence
we have considered $Y_{\Delta B}$ as the total baryon charge instead
of matrix in the baryon flavor space. The baryon flavor effect will be considered elsewhere. Hence, we will parametrize the
transitions across $T_{x}$ due to quark interactions, i.e. with $x\neq\left\{ e,\mu,\tau\right\} $,
as some exponential functions that we will discuss in the next section. Ignoring baryon flavor effect, let us define the charge matrix
%%%
\begin{eqnarray}
Y_{\widetilde{\Delta}} & \equiv & \frac{1}{3}Y_{\Delta B}I_{3\times3}-Y_{\Delta\ell},\label{eq:tildeDelta_matrix}
\end{eqnarray}
%%%
which transforms like $Y_{\Delta\ell}$ as in eq. \eqref{eq:transformation_leptonic_charges} under flavor rotations \eqref{eq:flavor_rotations}. From eqs. (\ref{eq:BE_Yell}) and
(\ref{eq:BE_YB}), we obtain the Boltzmann equation for $Y_{\widetilde{\Delta}}$
as follows
%%%
\begin{eqnarray}
s{\cal H}z\frac{dY_{\widetilde{\Delta}}}{dz} & = & \frac{\gamma_{E}}{2Y^{{\rm nor}}}\left\{ y_{E}^{\dagger}y_{E},\frac{Y_{\Delta\ell}}{g_{\ell}\zeta_{\ell}}\right\} -\frac{\gamma_{E}}{Y^{{\rm nor}}}y_{E}^{\dagger}y_{E}\frac{Y_{\Delta H}}{g_{H}\zeta_{H}}-\frac{\gamma_{E}}{Y^{{\rm nor}}}y_{E}^{\dagger}\frac{Y_{\Delta E}}{g_{E}\zeta_{E}}y_{E}.\label{eq:BE_YtildeDelta}
\end{eqnarray}
%%%
Now we only need to solve (\ref{eq:BE_YE}) and (\ref{eq:BE_YtildeDelta})
treating $Y_{\widetilde{\Delta}}$ and $Y_{\Delta E}$ as the only
independent variables. 

One could have defined the $B/3-L_{a}$ charge matrix 
%%%
\begin{eqnarray}
Y_{\Delta} & \equiv & \frac{1}{3}Y_{\Delta B}I_{3\times3}-Y_{\Delta\ell}-Y_{\Delta E},\label{eq:B-L_matrix}
\end{eqnarray}
%%%
where one would have to keep in mind that $Y_{\Delta\ell}$ and $Y_{\Delta E}$
transform differently as in eq. (\ref{eq:transformation_leptonic_charges}).
Clearly, the physics will remain the same but in order to avoid remembering the different transformations within $Y_{\Delta}$, we will resort
to using $Y_{\widetilde{\Delta}}$. Nevertheless, it is instructive
to look at the Boltzmann equation for $Y_{\Delta}$ in the flavor
basis where we can construct from eqs. (\ref{eq:BE_Yell}), (\ref{eq:BE_YE}),
and (\ref{eq:BE_YB}) as follows:
%%%
\begin{eqnarray}
s{\cal H}z\frac{dY_{\Delta}}{dz} & = & \frac{\gamma_{E}}{2Y^{{\rm nor}}}\left\{ \hat{y}_{E}^{2},\frac{Y_{\Delta\ell}}{g_{\ell}\zeta_{\ell}}\right\} -\frac{\gamma_{E}}{Y^{{\rm nor}}}\hat{y}_{E}\frac{Y_{\Delta E}}{g_{E}\zeta_{E}}\hat{y}_{E}\nonumber \\
 &  & +\frac{\gamma_{E}}{2Y^{{\rm nor}}}\left\{ \hat{y}_{E}^{2},\frac{Y_{\Delta E}}{g_{E}\zeta_{E}}\right\} -\frac{\gamma_{E}}{Y^{{\rm nor}}}\hat{y}_{E}\frac{Y_{\Delta\ell}}{g_{\ell}\zeta_{\ell}}\hat{y}_{E}\nonumber \\
 & = & -\frac{\gamma_{E}}{2Y^{{\rm nor}}}\left[\hat{y}_{E},\left[\hat{y}_{E},\frac{Y_{\Delta\ell}}{g_{\ell}\zeta_{\ell}}+\frac{Y_{\Delta E}}{g_{E}\zeta_{E}}\right]\right].\label{eq:BE_YDelta_flavor_basis}
\end{eqnarray}
%%%
In the last step above, we have defined the commutator $\left[A,B\right]\equiv AB-BA$.
One can easily check that in the flavor basis the double anticommutator
term projects out only the off-diagonal entries of $Y_{\Delta\ell}$ and
$Y_{\Delta E}$. Hence ${\rm Tr}Y_{\Delta}$ remains a constant as
it should be since the SM interactions do not break $B-L$. Clearly,
the same conclusion holds also in any other basis. While it is not
necessary to work in flavor basis, it makes the interpretation easier
since in the flavor basis, one can identify the diagonal elements
of $Y_{\Delta}$ as the flavor charges $Y_{\Delta(B/3-L_{e})}$, $Y_{\Delta\Delta(B/3-L_{\mu})}$
and $Y_{\Delta(B/3-L_{\tau})}$. In eq. (\ref{eq:BE_YDelta_flavor_basis}),
it is apparent that for a consistent description of evolution of lepton
flavor charges which is basis independent both $Y_{\Delta\ell}$ and $Y_{\Delta E}$ need to
be described by density matrices: if off-diagonal terms of $Y_{\Delta\ell}$
are induced, off-diagonal terms for $Y_{\Delta E}$ will be induced
as well and vice versa.

In the rest of the work, we will use eqs. \eqref{eq:BE_YE} and \eqref{eq:BE_YtildeDelta},
which are valid in any basis. Including new physics interactions that
generate either $Y_{\Delta E}$ and/or $Y_{\widetilde{\Delta}}$ in the
two Boltzmann equations, from eq. (\ref{eq:B-L_to_B}), the final
baryon asymmetry will be frozen at $T_{B-}$ to be
%%%
\begin{eqnarray}
Y_{\Delta B}\left(T_{B-}\right) & = & \left.\frac{30}{97}\left({\rm Tr}Y_{\widetilde{\Delta}}-{\rm Tr}Y_{\Delta E}\right)\right|_{T=T_{B-}}.
\end{eqnarray}
%%%

Next, we will write down the relations between $Y_{\Delta\ell}$ and
$Y_{\Delta H}$ in terms of $Y_{\widetilde{\Delta}}$ and $Y_{\Delta E}$
for the SM and the SM augmented with arbitrary scalar fields carrying
nonzero hypercharges.

\subsection{Standard Model\label{subsec:SM}}

With the SM field content, from eq. (\ref{eq:YDi_YDx}), we obtain\footnote{The number asymmetries of quark fields in term of $Y_{\widetilde{\Delta}}$
and $Y_{\Delta E}$ are collected in Appendix \ref{app:quark_relations}.}
%%%
\begin{eqnarray}
\left(Y_{\Delta\ell}\right)_{\alpha\alpha} & = & \frac{2}{15}c_{B}{\rm Tr}Y_{\widetilde{\Delta}}-\left(Y_{\widetilde{\Delta}}\right)_{\alpha\alpha},\label{eq:Yell_Ycharges}\\
Y_{\Delta H} & = & -c_{H}\left({\rm Tr}Y_{\widetilde{\Delta}}-2{\rm Tr}Y_{\Delta E}\right),\label{eq:YH_Ycharges}
\end{eqnarray}
%%%
where $c_{B}$ and $c_{H}$ are coefficients which vary with temperature. In obtaining the expressions above, we have assumed all effective charges in eq. \eqref{eq:SM_global_symmetries}, except $(Y_{\widetilde \Delta})_{\alpha\alpha}$ and $(Y_{\Delta E})_{\alpha\alpha}$, to be zero.
Comparing with eq. (\ref{eq:tildeDelta_matrix}), one recognizes $\frac{B}{3}=\frac{2}{15}c_{B}{\rm Tr}Y_{\widetilde{\Delta}}$.
The relations above are completely general and capture \emph{all}
the spectator effects in the SM. At $T>T_{B}\sim2\times10^{12}$ GeV
when the EW sphaleron interaction is out of equilibrium, we have $c_{B}=0$
while at $T<T_{B}$ when the baryon number is no longer conserved,
we have $c_{B}=1$. This shows that an asymmetry in the lepton sector is being shared with the baryon sector and vice versa at $T<T_{B}$.
To capture this effect in a continuous manner, one should consider $Y_{\Delta\ell_{a}}=\frac{B}{3}-Y_{\widetilde{\Delta}_{a}}$ and include
the Boltzmann equation for $Y_{\Delta B}$ in eq. (\ref{eq:BE_YB})
and then solve for $c_{B}\left(T\right)$. To within percent-level
precision, one can use the fitting function\footnote{One can also use a theta function $c_{B}\left(T\right)=\theta\left(T-T_{B}\right)$
keeping in mind that the effect can be of the order of 1 if leptogenesis
happens around $T_{B}$.}
\begin{eqnarray}
c_{B}\left(T\right) & = & 1-e^{-\frac{T_{B}}{T}},
\end{eqnarray}
%%%
where $T_{B}=2.3\times10^{3}$ GeV. In Appendix \ref{app:trans_temps},
we discuss how to determine a precise value of $T_{B}$.

The rest of the spectator effects pertaining to quark sector are encapsulated
in the coefficient $c_{H}$ with
%%%
\begin{eqnarray}
c_{H}\left(T\right) & = & \begin{cases}
1 & T>T_{t}\\
\frac{2}{3} & T_{u}<T<T_{t}\\
\frac{14}{23} & T_{u-b}<T<T_{u}\\
\frac{2}{5} & T_{u-c}<T<T_{u-b}\\
\frac{4}{13} & T_{B_{3}-B_{2}}<T<T_{u-c}\\
\frac{3}{10} & T_{u-s}<T<T_{B_{3}-B_{2}}\\
\frac{1}{4} & T_{u-d}<T<T_{u-s}\\
\frac{2}{11} & T<T_{u-d}
\end{cases}\label{eq:cH}.
\end{eqnarray}
%%%
In the equation above, we can see explicitly that the asymmetry carried by the Higgs is diluted as more
charges come into equilibrium. Since the transitions due to the rate
$\Gamma\propto T$ as compared to the Hubble rate $H\propto T^{2}$
always have an exponential behavior, one can parametrize the transitions
with the following function
%%%
\begin{eqnarray}
c_{H}\left(T\right) & = & \left(\frac{2}{3}+\frac{1}{3}e^{-\frac{T_{t}}{T}}\right)-\left(\frac{2}{3}-\frac{14}{23}\right)\left(1-e^{-\frac{T_{u}}{T}}\right)-\left(\frac{14}{23}-\frac{2}{5}\right)\left(1-e^{-\frac{T_{u-b}}{T}}\right)\nonumber \\
 &  & -\left(\frac{2}{5}-\frac{4}{13}\right)\left(1-e^{-\frac{T_{u-c}}{T}}\right)-\left(\frac{4}{13}-\frac{3}{10}\right)\left(1-e^{-\frac{T_{B_{3}-B_{2}}}{T}}\right)\nonumber \\
 &  & -\left(\frac{3}{10}-\frac{1}{4}\right)\left(1-e^{-\frac{T_{u-s}}{T}}\right)-\left(\frac{1}{4}-\frac{2}{11}\right)\left(1-e^{-\frac{T_{u-d}}{T}}\right).
\end{eqnarray}
%%%
For the purpose of this work, we use the transition temperatures as shown in eq. (\ref{eq:transition_temperatures}). Precise determination
of the transition temperatures can be carried out following the procedure
shown in Appendix \ref{app:trans_temps}. 

From the definition of $Y_{\widetilde{\Delta}}$ in eq. (\ref{eq:tildeDelta_matrix}),
the off-diagonal terms $\alpha\neq\beta$ are
%%%
\begin{eqnarray}
\left(Y_{\widetilde{\Delta}}\right)_{\alpha\beta} & = & -\left(Y_{\Delta\ell}\right)_{\alpha\beta}.
\end{eqnarray}
%%%
Hence we can rewrite the matrix $Y_{\Delta\ell}$ as
%%%
\begin{eqnarray}
Y_{\Delta\ell} & = & \frac{2}{15}c_{B}I_{3\times3}{\rm Tr}Y_{\widetilde{\Delta}}-Y_{\widetilde{\Delta}}.\label{eq:Yellmatrix_Ycharges}
\end{eqnarray}
%%%

\subsection{Standard Model with additional scalar fields\label{subsec:SM_scalars}}

If one introduces additional scalar fields $\phi_{i}$ with hypercharge
$q_{\phi_{i}}^{Y}$ to the system, eq. (\ref{eq:Yell_Ycharges}) remains
the same, while eq. (\ref{eq:YH_Ycharges}) changes to
%%%
\begin{eqnarray}
Y_{\Delta H} & = & -c_{H}\left({\rm Tr}Y_{\widetilde{\Delta}}-2{\rm Tr}Y_{\Delta E}+2\sum_{i}q_{\phi_{i}}^{Y}Y_{\Delta\phi_{i}}\right),\label{eq:YH_Ycharges_general}
\end{eqnarray}
%%%
where $Y_{\Delta\phi_{i}}$ defined in eq. (\ref{eq:Y_mu_i}) takes
into account additional gauge multiplicity $g_{\phi_{i}}$ as well
as mass of $\phi_{i}$ in $\zeta_{\phi_{i}}$ (implicitly, we have
assumed $\phi_{i}$ to be in kinetic equilibrium but not necessarily
in chemical equilibrium). The relation above is general, independently
of whether $\phi_{i}$ are in chemical equilibrium or not. If some
of the $\phi_{i}$ do not achieve chemical equilibrium, one will have
effective $U(1)_{\phi_{i}}$ in which $Y_{\Delta\phi_{i}}$ remains
constant. Otherwise, the evolution of $Y_{\Delta\phi_{i}}$ will have
to be described by the corresponding Boltzmann equation. 

For instance, for type-II seesaw leptogenesis with a heavy triplet
Higgs ${\cal T}$ with hypercharge $q_{{\cal T}}^{Y}=1$, one can
apply eq. (\ref{eq:YH_Ycharges_general}) and obtain
%%%
\begin{eqnarray}
Y_{\Delta H} & = & -c_{H}\left({\rm Tr}Y_{\widetilde{\Delta}}-2{\rm Tr}Y_{\Delta E}+2Y_{\Delta{\cal T}}\right).\label{eq:YH_Ycharges_HiggsTriplets}
\end{eqnarray}
%%%

\section{Applications\label{sec:Applications}}

Now, we will apply the flavor-covariant Boltzmann equations \eqref{eq:BE_YE}
and \eqref{eq:BE_YtildeDelta} to some well-motivated leptogenesis
scenarios. One just needs the general expressions \eqref{eq:Yellmatrix_Ycharges}
and \eqref{eq:YH_Ycharges_general} to close the equations. Even for
leptogenesis models involving quarks, one can use the general relations
in Appendix \ref{app:quark_relations} (ignoring baryon flavor effect).
Hence, one no longer needs to solve for flavor matrices for a particular model and which hold only in a particular temperature regime as has been done, for example, in refs. \cite{Nardi:2006fx} and \cite{Lavignac:2015gpa}. In the first example, we will apply the formalism to type-I
leptogenesis, while in the second example, we will apply it to type-II
leptogenesis where flavor-covariant formalism is indispensable as
first pointed out in ref. \cite{Lavignac:2015gpa}. In particular,
we will demonstrate that the results obtained are independent of basis, showing that it is necessary to take into account flavor correlation in both $\ell$ and $E$. In other words, it is inconsistent
to consider flavor correlation only in $\ell$ or only in $E$.

\subsection{Type-I leptogenesis}

In the type-I seesaw model, the SM is extended by right-handed neutrinos
$N_{i}$ as
%%%
\begin{eqnarray}
-{\cal L} & \supset & \frac{1}{2}M_{i}\overline{N_{i}}N_{i}^{c}+y_{i\alpha}\overline{N_{i}}\ell_{\alpha}\epsilon H+{\rm H.c.},\label{eq:type_I}
\end{eqnarray}
%%%
where $M_{i}$ is the Majorana mass of $N_{i}$ and we will work in the arbitrary basis where $y_{E}$ is not necessarily diagonal. While two generations
of $N_{i}$ are already sufficient to explain neutrino oscillation
data, as an example, we will consider three generations $i=1,2,3$. 

After the EW symmetry breaking with $v\equiv\left\langle H\right\rangle =174$
GeV, the light neutrino mass matrix for $\left|y\right|v\ll M_{i}$
is 
%%%
\begin{eqnarray}
m_{\nu}^{{\rm I}} & = & -v^{2}y^{T}M^{-1}y,
\end{eqnarray}
%%%
where $M={\rm diag}\left(M_{1},M_{2},M_{3}\right)$. The mass matrix
can be diagonalized with $U_{\nu}^{T}m_{\nu}U_{\nu}=\hat{m}\equiv{\rm diag}\left(m_{1},m_{2},m_{3}\right)$
where $U_{{\rm PMNS}}=V_{E}U_{\nu}$ is identified with the leptonic
mixing matrix.

For type-I leptogenesis, an asymmetry is generated through the $CP$-violating decays $N_i \to \ell_\alpha H$. In addition to the Boltzmann equation for
$Y_{N_{i}}$,
%%%
\begin{equation}
s{\cal H}z\frac{dY_{N_{i}}}{dz}=-\gamma_{N_{i}}\left(\frac{Y_{N_{i}}}{Y_{N_{i}}^{{\rm eq}}}-1\right),
\end{equation}
%%%
where we have defined $z\equiv M_{1}/T$, we have to append to the
right-hand side of eq. (\ref{eq:BE_YtildeDelta}) a source and washout terms, respectively, given by \cite{Blanchet:2011xq}
%%%
\begin{eqnarray}
S^{{\rm I}} & \equiv & -\sum_{i}\epsilon_{i}\gamma_{N_{i}}\left(\frac{Y_{N_{i}}}{Y_{N_{i}}^{{\rm eq}}}-1\right),\\
W^{{\rm I}} & \equiv & \frac{1}{2}\sum_{i}\frac{\gamma_{N_{i}}}{Y^{{\rm nor}}}\left(\frac{1}{2}\left\{ P_{i},\frac{Y_{\Delta\ell}}{g_{\ell}\zeta_{\ell}}\right\} +P_{i}\frac{Y_{\Delta H}}{g_{H}\zeta_{H}}\right),
\end{eqnarray}
%%%
where to close the equations we apply eqs. \eqref{eq:YH_Ycharges}
and \eqref{eq:Yellmatrix_Ycharges}. Assuming the Maxwell-Boltzmann
distribution for $N_{i}$, we have $Y_{N_{i}}^{{\rm eq}}=\frac{45}{2\pi^{4}g_{\star}}\frac{M_{i}^{2}}{T^{2}}{\cal K}_{2}\left(\frac{M_{i}}{T}\right)$
with ${\cal K}_{n}\left(x\right)$ the modified Bessel function of
the second kind of order $n$ and the decay reaction density $\gamma_{N_{i}}$
is given by
%%%
\begin{eqnarray}
\gamma_{N_{i}} & = & sY_{N_{i}}^{{\rm eq}}\Gamma_{N_{i}}\frac{{\cal K}_{1}\left(M_{i}/T\right)}{{\cal K}_{2}\left(M_{i}/T\right)},
\end{eqnarray}
%%%
with $\Gamma_{N_{i}}=\frac{\left(yy^{\dagger}\right)_{ii}M_{i}}{8\pi}$
the total decay width of $N_{i}$.\footnote{Here we consider only decay and inverse decay. We have ignored the helicities of $N_{i}$ and scattering processes which will be relevant
for leptogenesis in the weak washout regime $\Gamma_{N_{i}}/{\cal H}\left(T=M_{i}\right)\ll1$
since in this case, the physics at $T\gg M_{i}$ will play a relevant role \cite{Garbrecht:2019zaa}. We have also assumed $N_i$ to be well-separated states $|M_i - M_j| \gg \Gamma_{N_{i,j}}$ such that the effect of $N_i$ oscillations is not relevant. Otherwise, one should use the flavor-covariant formalism which also includes the flavor of $N_i$ \cite{BhupalDev:2014oar}.}

The matrix of $CP$-violation parameter $\epsilon_{i}$ and flavor rotation
matrix $P_{i}$ are, respectively, given by \cite{Blanchet:2011xq}
%%%
\begin{eqnarray}
\left(\epsilon_{i}\right)_{\alpha\beta} & = & \frac{1}{16\pi}\frac{i}{\left(yy^{\dagger}\right)_{ii}}\sum_{j\neq i}\left[\left(yy^{\dagger}\right)_{ji}y_{j\beta}y_{i\alpha}^{*}-\left(yy^{\dagger}\right)_{ij}y_{i\beta}y_{j\alpha}^{*}\right]g\left(\frac{M_{j}^{2}}{M_{i}^{2}}\right)\nonumber \\
 &  & +\frac{1}{16\pi}\frac{i}{\left(yy^{\dagger}\right)_{ii}}\sum_{j\neq i}\left[\left(yy^{\dagger}\right)_{ij}y_{j\beta}y_{i\alpha}^{*}-\left(yy^{\dagger}\right)_{ji}y_{i\beta}y_{j\alpha}^{*}\right]\frac{M_{i}^{2}}{M_{i}^{2}-M_{j}^{2}},\label{eq:CP_typeI}\\
P_{i} & = & \frac{1}{\left(yy^{\dagger}\right)_{ii}}\left(\begin{array}{ccc}
\left|y_{ie}\right|^{2} & y_{ie}^{*}y_{i\mu} & y_{ie}^{*}y_{i\tau}\\
y_{ie}y_{i\mu}^{*} & \left|y_{i\mu}\right|^{2} & y_{i\mu}^{*}y_{i\tau}\\
y_{ie}y_{i\tau}^{*} & y_{i\mu}y_{i\tau}^{*} & \left|y_{i\tau}\right|^{2}
\end{array}\right).
\end{eqnarray}
%%%
Under flavor rotations \eqref{eq:flavor_rotations} and \eqref{eq:transformation_leptonic_charges},
we have
%%%
\begin{eqnarray}
\epsilon_{i} & \to & V\epsilon_{i}V^{\dagger},\;\;\;\;\;P_{i}\to VP_{i}V^{\dagger},
\end{eqnarray}
%%%
and the whole Boltzmann equation for $Y_{\widetilde \Delta}$ remains flavor covariant as required.

For illustration, we choose the best-fit point from ref. \cite{Dueck:2013gca}
for the SO(10) model with Higgs content $10_{H}+\overline{126}_{H}$ for
the Yukawa sector with no-RGE 
%%%
{\small
\begin{eqnarray}
y & = & \left(\begin{array}{ccc}
\left(2.508-1.101i\right)\times10^{-4} & \left(1.224-5.313i\right)\times10^{-4} & \left(-1.988+0.646i\right)\times10^{-2}\\
\left(1.893+0.0359i\right)\times10^{-3} & \left(-2.100+20.365i\right)\times10^{-3} & \left(-8.560+1.384i\right)\times10^{-2}\\
\left(1.446-9.365i\right)\times10^{-3} & \left(2.217+1.373i\right)\times10^{-2} & 0.1356+0.4602i
\end{array}\right),\\
y_{E} & = & \left(\begin{array}{ccc}
\left(1.0077+1.0449i\right)\times10^{-5} & \left(-3.8245+0.0226i\right)\times10^{-5} & \left(-3.2332-1.8088i\right)\times10^{-4}\\
\left(-3.8245+0.0226i\right)\times10^{-5} & \left(5.2064-2.2026i\right)\times10^{-4} & \left(8.0184-7.4693i\right)\times10^{-4}\\
\left(-3.2332-1.8088i\right)\times10^{-4} & \left(8.0184-7.4693i\right)\times10^{-4} & \left(8.5102+4.4337i\right)\times10^{-3}
\end{array}\right),\\
M & = & \left\{ 1.445\times10^{10},7.244\times10^{11},5.663\times10^{12}\right\} \,{\rm GeV}.
\end{eqnarray}
}
%%%
We will solve the Boltzmann equations in the original basis (as above) and in the flavor basis where $y_{E}$ is diagonalized through a flavor rotation as in eq. (\ref{eq:flavor_basis}) assuming zero initial abundance $Y_{N_{i}}\left(z_{i}\right)=0$ with $z_i = 10^{-4}$. 

\begin{figure}[H]
	\begin{centering}
		\includegraphics[scale=0.3]{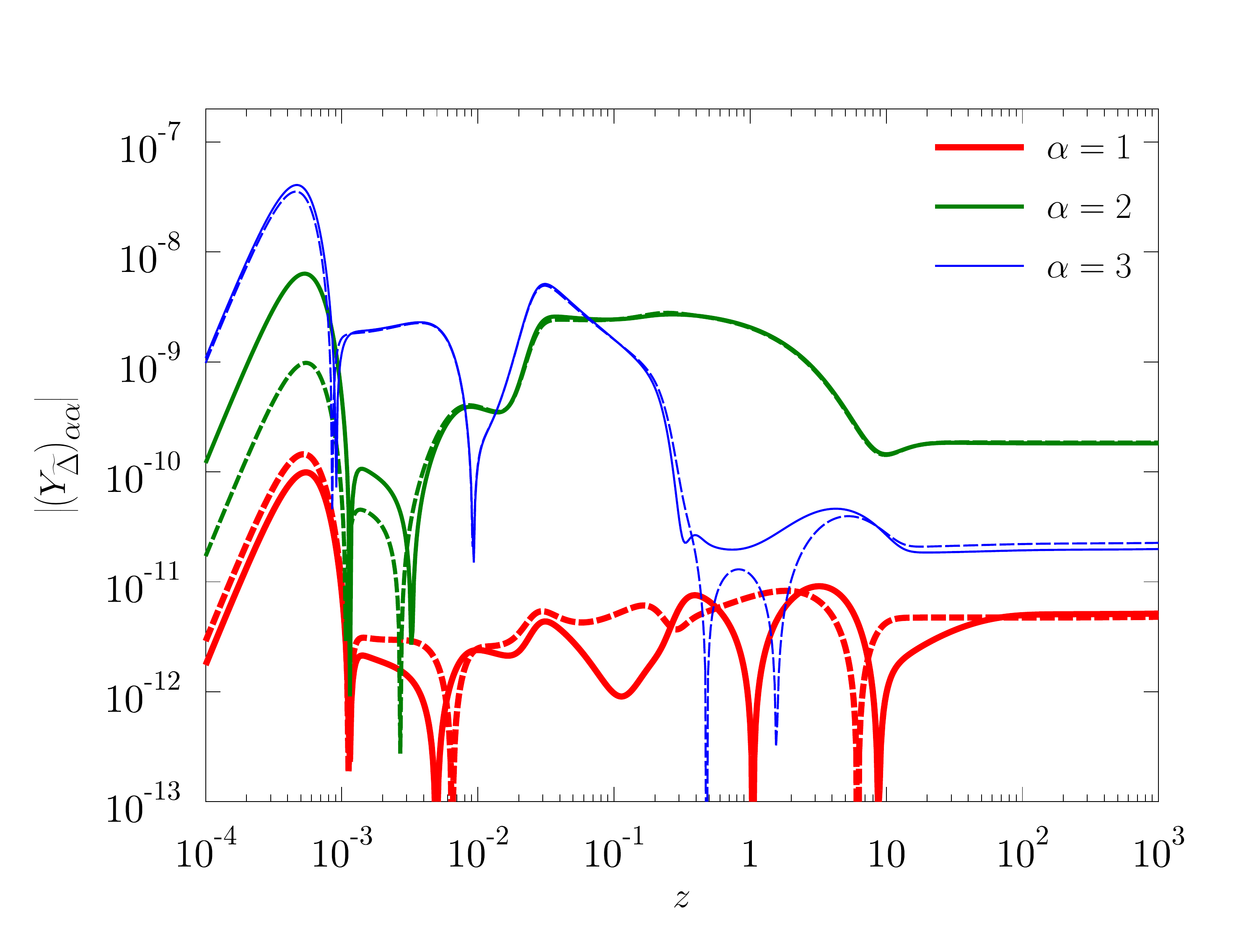}\includegraphics[scale=0.3]{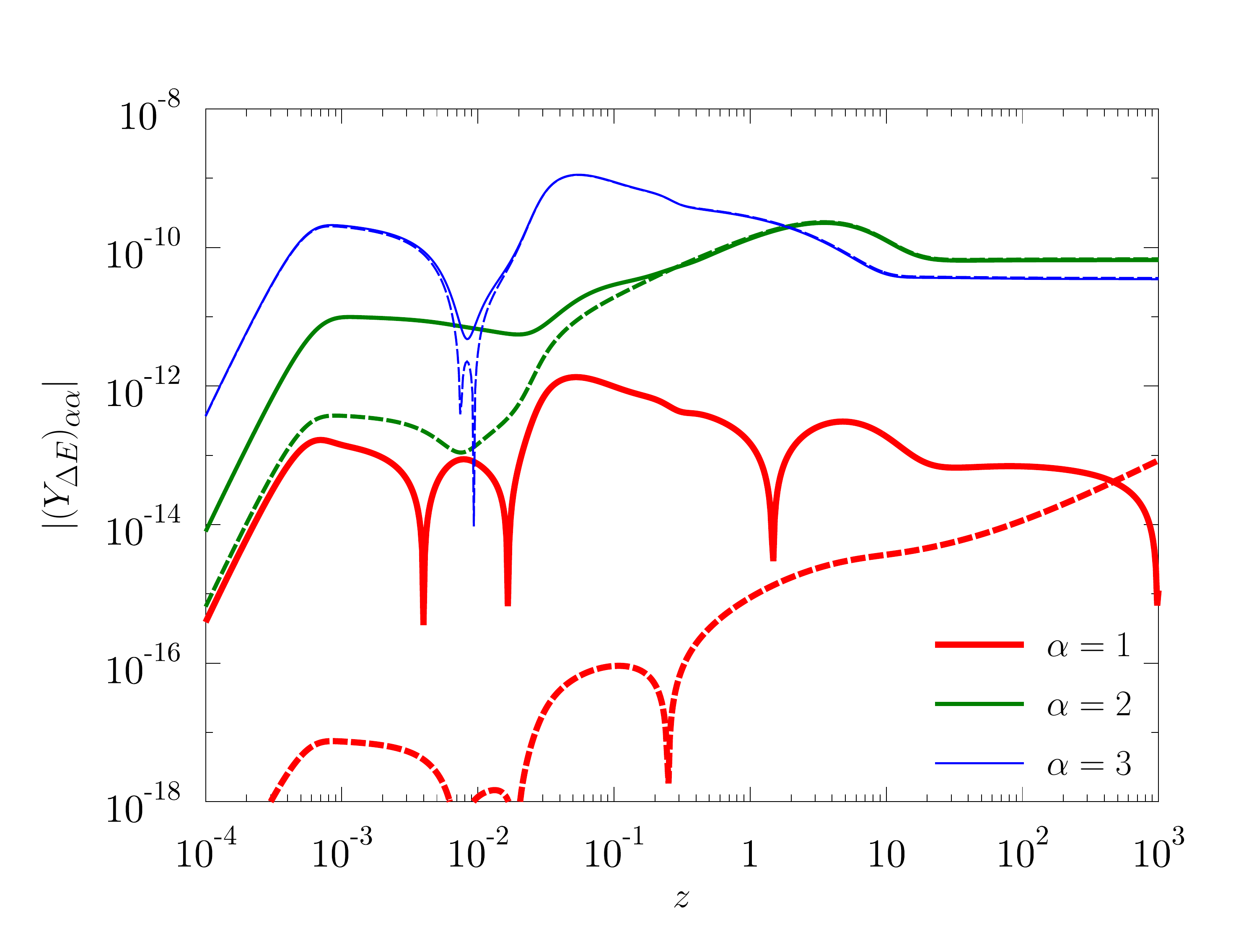}
		\par\end{centering}
	\begin{centering}
		\includegraphics[scale=0.3]{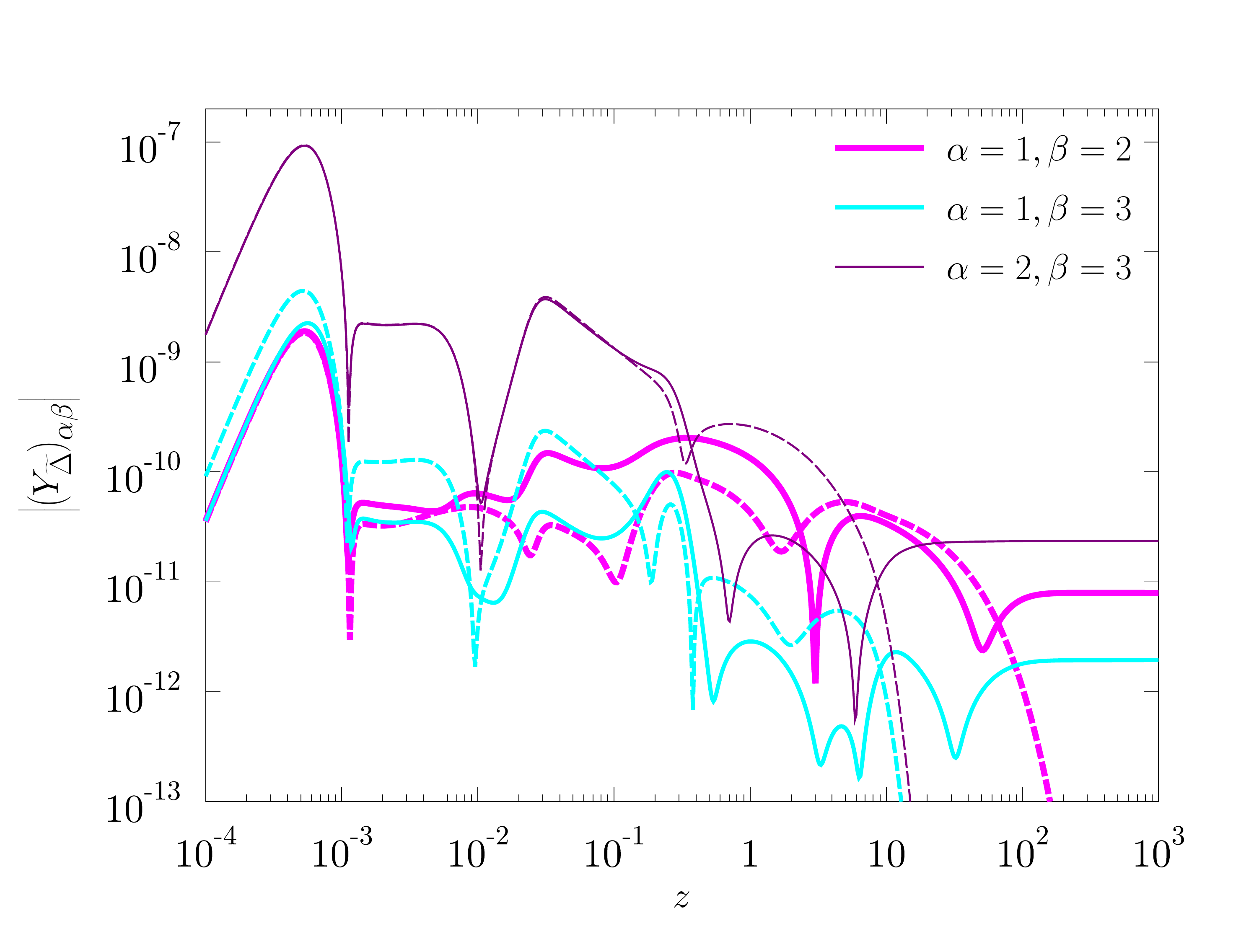}\includegraphics[scale=0.3]{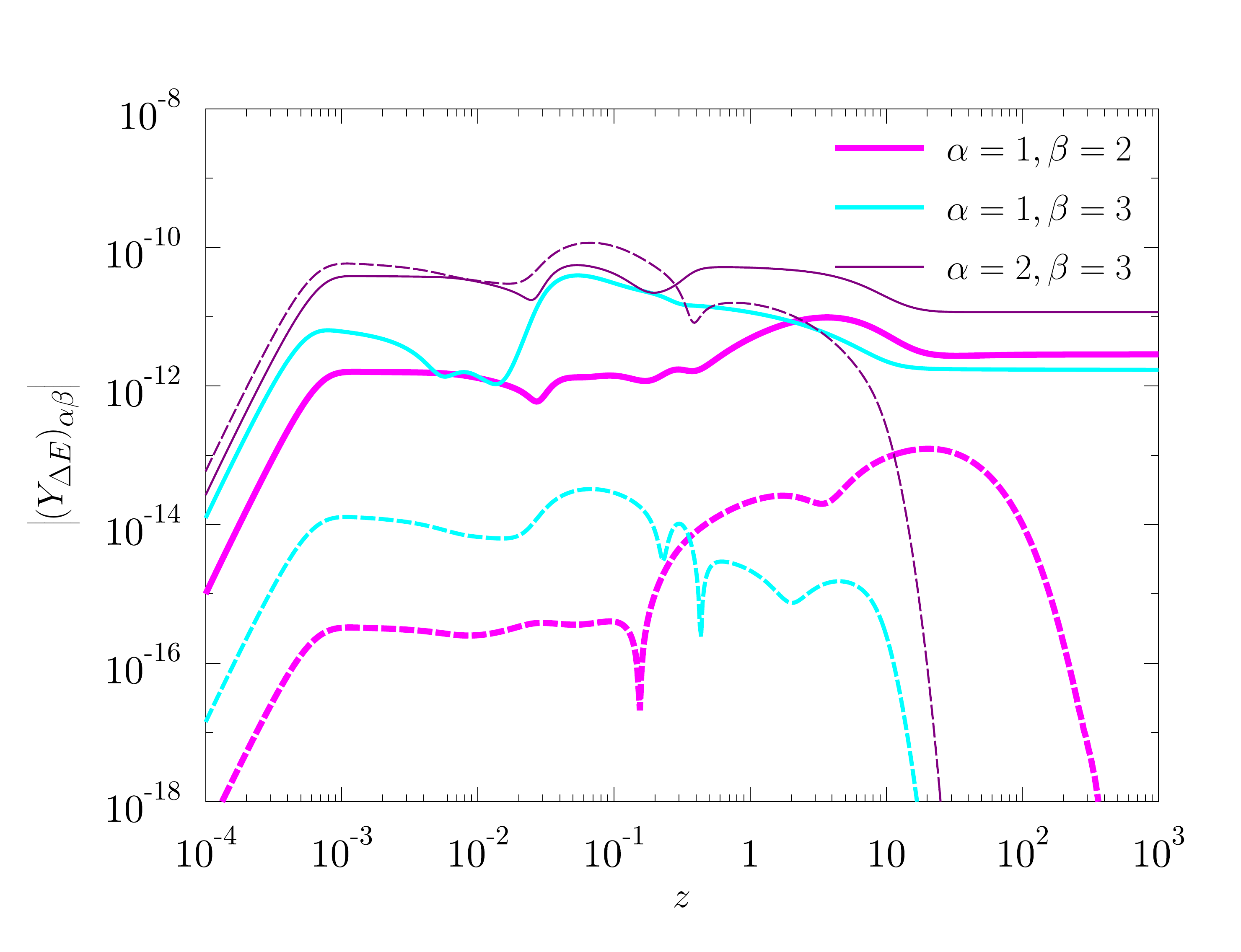}
		\par\end{centering}
	\caption{Numerical solutions for type-I leptogenesis. In the top row, we plot
		the diagonal elements of $\left|Y_{\widetilde{\Delta}}\right|$ and
		$\left|Y_{\Delta E}\right|$ in the two different bases: nonflavor
		basis $y_{E}$ (solid curves) and flavor basis $\hat{y}_{E}$ (dashed
		curves). In the bottom row, we plot the off-diagonal elements of $\left|Y_{\widetilde{\Delta}}\right|$
		and $\left|Y_{\Delta E}\right|$ in the $y_{E}$ (solid curves) and $\hat{y}_{E}$ (dashed curves) bases. Colors (thickness) denote different matrix elements as indicated in the plots. See the
		text for further discussions. \label{fig:type-I_leptogenesis}}
	
\end{figure}

\begin{figure}[H]
	\begin{centering}
		\includegraphics[scale=0.3]{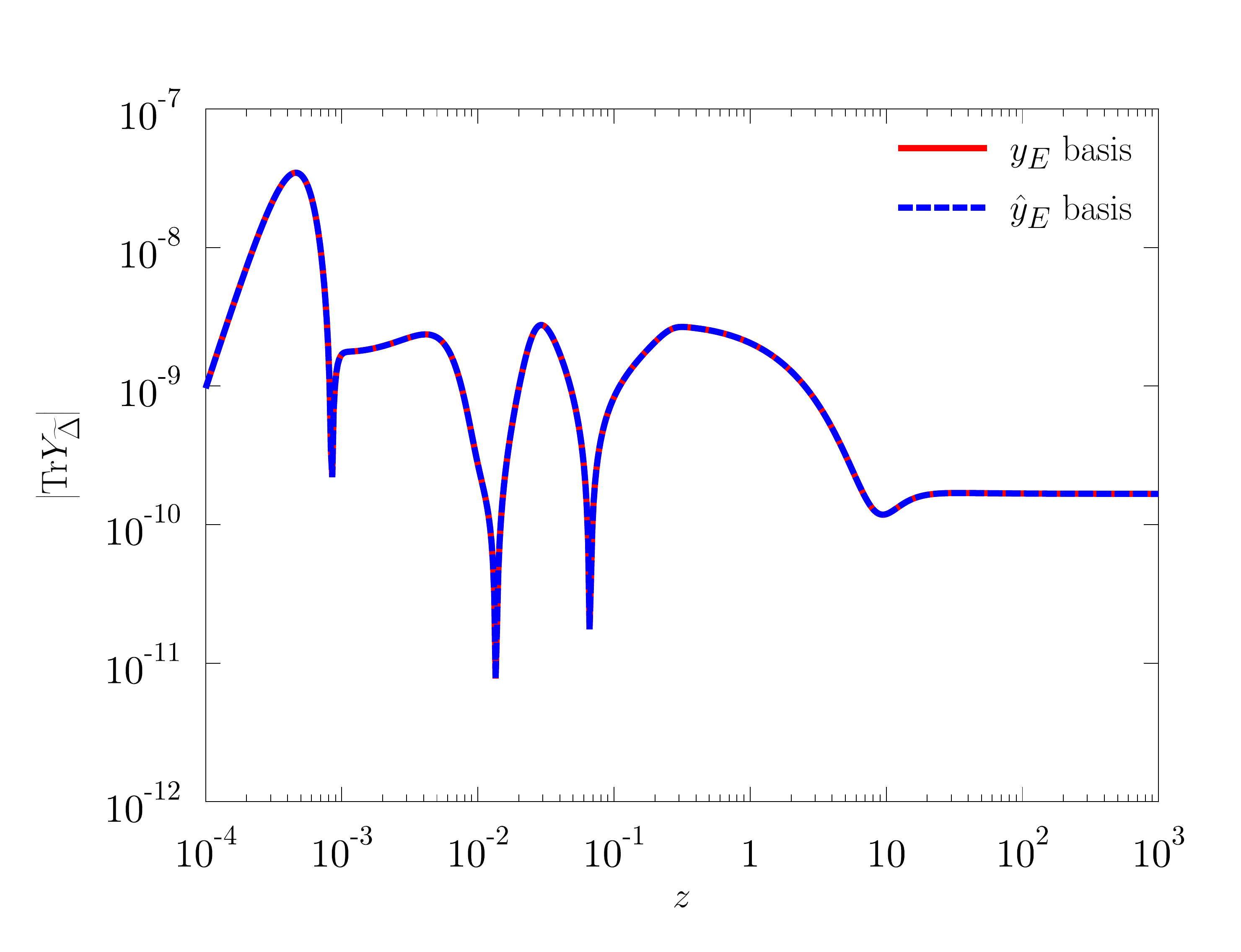}\includegraphics[scale=0.3]{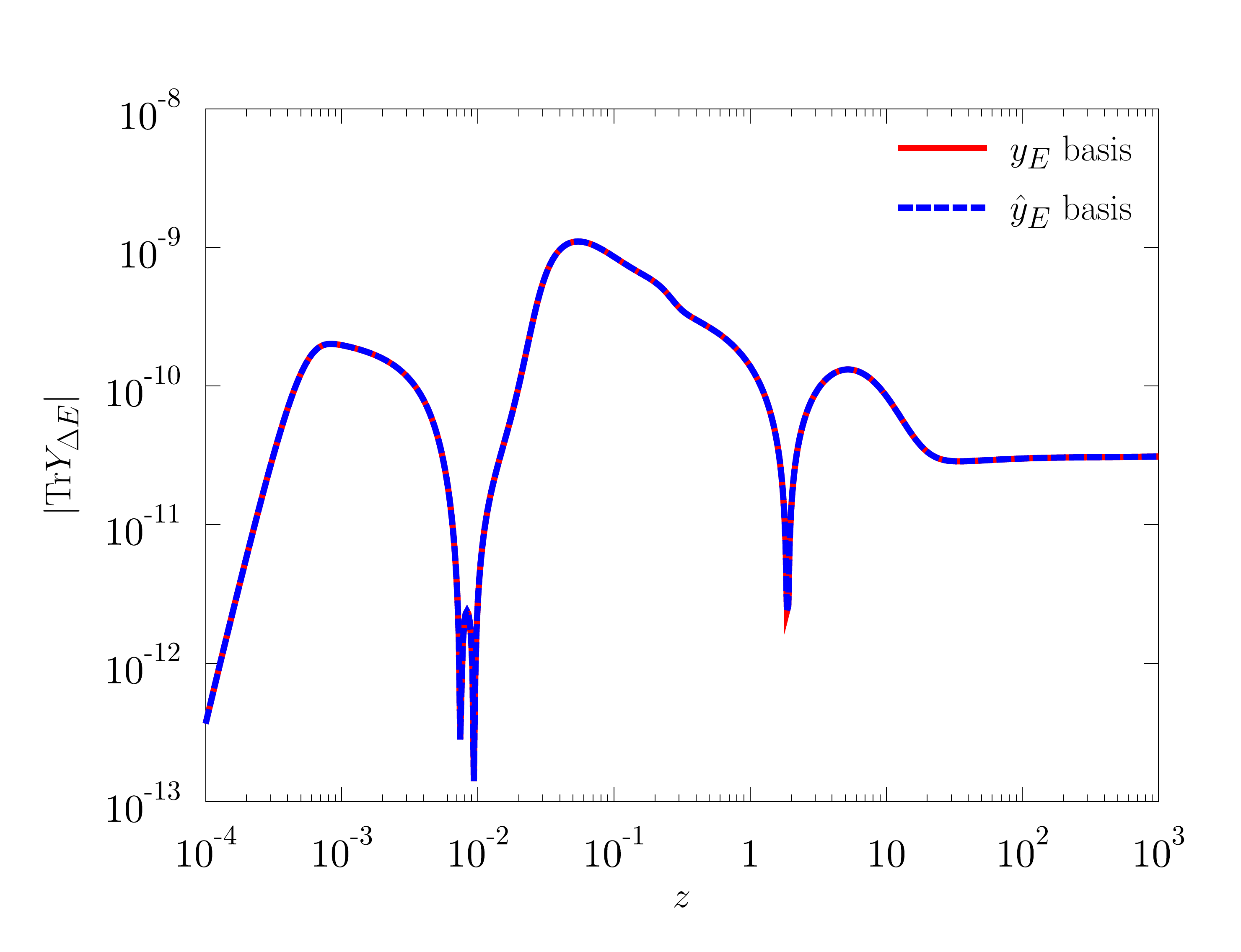}
		\par\end{centering}
	\begin{centering}
		\includegraphics[scale=0.3]{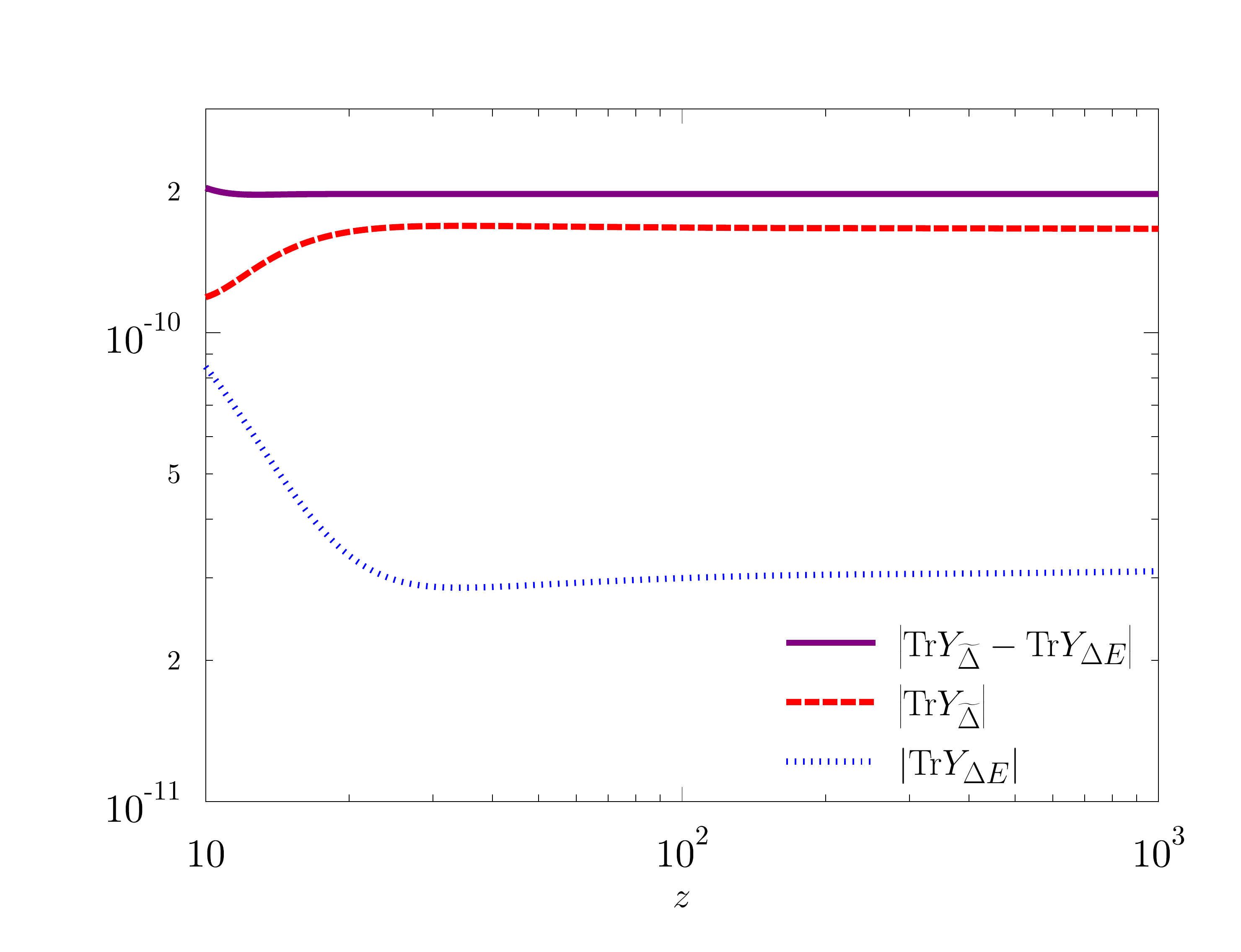}
		\par\end{centering}
	\caption{Numerical solutions for type-I leptogenesis. In the top row, we plot
		$\left|{\rm Tr}Y_{\widetilde{\Delta}}\right|$ and $\left|{\rm Tr}Y_{\Delta E}\right|$
		in the two different bases: nonflavor basis $y_{E}$ (red solid curve)
		and flavor basis $\hat{y}_{E}$ (blue dashed curve). In the bottom
		plot, we plot $\left|{\rm Tr}Y_{\widetilde{\Delta}}-{\rm Tr}Y_{\Delta E}\right|$
		(purple solid curve), $\left|{\rm Tr}Y_{\widetilde{\Delta}}\right|$
		(red dashed curve) and $\left|{\rm Tr}Y_{\Delta E}\right|$ (blue
		dotted curve) for $z>10$. See the text for further discussions. \label{fig:type-I_leptogenesis-2}}
\end{figure}

In Figure \ref{fig:type-I_leptogenesis}, we show the numerical solutions
comparing the results in the nonflavor basis $y_{E}$ (solid curves)
and in the flavor basis $\hat{y}_{E}$ (dashed curves). In the top
row, we show the diagonal elements of $\left|Y_{\widetilde{\Delta}}\right|$
and $\left|Y_{\Delta E}\right|$, while in the bottom row, we show their off-diagonal
elements (they are Hermitian matrices).
Here, we see that independent of basis, once off-diagonal elements
of $Y_{\widetilde{\Delta}}$ develop from leptogenesis, unavoidably,
off-diagonal elements of $Y_{\Delta E}$ will be induced as well. In
the flavor basis $\hat{y}_{E}$, the off-diagonal elements start to
become suppressed at various temperatures as the charged lepton Yukawa
interactions subsequently get into thermal equilibrium and finally
at $z\gtrsim100$, $\left(Y_{\widetilde{\Delta}}\right)_{12}$ and
$\left(Y_{\Delta E}\right)_{12}$ start to become suppressed, indicating
a transition to the three-flavor regime.

In Figure \ref{fig:type-I_leptogenesis-2}, top row, we plot $\left|{\rm Tr}Y_{\widetilde{\Delta}}\right|$
and $\left|{\rm Tr}Y_{\Delta E}\right|$ in the two different bases:
nonflavor basis $y_{E}$ (red solid curve) and flavor basis $\hat{y}_{E}$
(blue dashed curve). Reassuringly, ${\rm Tr}Y_{\widetilde{\Delta}}$
and ${\rm Tr}Y_{\Delta E}$ are basis independent, although the entries
of $Y_{\widetilde{\Delta}}$ and $Y_{\Delta E}$ differ among the
two bases by the flavor rotations as in eq. (\ref{eq:transformation_leptonic_charges})
with $V=V_{E}$ and $U=V_{E}^{*}$ since $y_{E}$ is symmetric. Clearly,
the physics is invariant under basis transformation, and the benefit
of the flavor basis is to help us to interpret the results. For instance,
we can read the diagonal entries of $Y_{\widetilde{\Delta}}$ and
$Y_{\Delta E}$ in the $\hat y_E$ basis as the flavor charges in the $e,\mu,\tau$ (red, blue
and green dashed curves in the top row of the Figure \ref{fig:type-I_leptogenesis})
and also deduce when the system transits to a different flavor regime from the suppression
of off-diagonal entries. In the bottom plot of Figure \ref{fig:type-I_leptogenesis-2},
we see that, while $\left|Y_{\Delta(B-L)}\right|=\left|{\rm Tr}Y_{\widetilde{\Delta}}-{\rm Tr}Y_{\Delta E}\right|$
is conserved at the end of leptogenesis $z\gtrsim10$, $\left|{\rm Tr}Y_{\widetilde{\Delta}}\right|$
and $\left|{\rm Tr}Y_{\Delta E}\right|$, not being conserved charges,
continue to evolve.
For a final remark, the final baryon asymmetry produced in this example is $Y_{\Delta B}\left(T_{B-}\right)=6.1\times10^{-11}$,
consistent in sign but smaller than the observed value by about 30\%.\footnote{This can be compared with ref. \cite{Fong:2014gea}, which also obtained
a final baryon asymmetry, which is of the right sign but a factor of
a few smaller than the observed baryon asymmetry. Besides the improved
treatment discussed in the work, we also correct the wrong basis used
in ref. \cite{Fong:2014gea}.}

\subsection{Type-II leptogenesis}

In the type-II seesaw model, the SM is extended by a massive triplet
scalar ${\cal T}$ under $SU(2)_{L}$ with hypercharge $q_{{\cal T}}^{Y}=1$
as
%%%
\begin{eqnarray}
-{\cal L} & \supset & M_{{\cal T}}^{2}{\rm Tr}\left({\cal T}^{\dagger}{\cal T}\right)+\frac{1}{2}\left(f_{\alpha\beta}\overline{\ell_{\alpha}^{c}}\epsilon{\cal T}\ell_{\beta}+\mu H^{T}\epsilon{\cal T}^{\dagger}H+{\rm H.c.}\right),\label{eq:type_II}
\end{eqnarray}
%%%
where 
%%%
\begin{eqnarray}
{\cal T} & = & \left(\begin{array}{cc}
\frac{1}{\sqrt{2}}{\cal T}^{+} & {\cal T}^{++}\\
{\cal T}^{0} & -\frac{1}{\sqrt{2}}{\cal T}^{+}
\end{array}\right).
\end{eqnarray}
%%%
Since ${\cal T}$ couples to two lepton doublets which in general do not align in flavor space, one needs to describe them with density matrix as first pointed out in ref. \cite{Lavignac:2015gpa}. 

The $CP$ violation in the decays of ${\cal T}^{\dagger}\to\ell_{\alpha}\ell_{\beta}$
and ${\cal T}\to HH$ can arise at one-loop level from the contribution
of heavier particles of mass scale $\Lambda\gg M_{T}$, which generate
the Weinberg operator below $\Lambda$,
%%%
\begin{eqnarray}
{\cal L}_{{\rm eff}} & = & \frac{1}{4}\frac{\kappa_{\alpha\beta}}{\Lambda}\overline{\ell_{\alpha}^{c}}\epsilon H\,H^{T} \epsilon \ell_{\beta}+{\rm H.c.}.
\end{eqnarray}
%%%
After the EW symmetry breaking, the light neutrino mass receives contributions
from integrating out the scalar triplet $T$ as well as the Weinberg
operator as
%%%
\begin{eqnarray}
m_{\nu}^{{\rm II}} & = & m_{{\cal T}}+m_{{\cal H}},
\end{eqnarray}
%%%
where 
%%%
\begin{eqnarray}
m_{{\cal T}} & \equiv & \frac{1}{2}\mu f\frac{v^{2}}{M_{T}^{2}},\\
m_{{\cal H}} & \equiv & \frac{1}{2}\kappa\frac{v^{2}}{\Lambda}.
\end{eqnarray}
%%%

In the following, we will utilize the interaction terms derived in ref. \cite{Lavignac:2015gpa} but include only decay, inverse decay
and gauge scattering processes (other scattering effects are negligible
in the parameter space we will consider below). The Boltzmann equations
to describe the evolution of $Y_{\Sigma{\cal T}}\equiv Y_{{\cal T}}+Y_{{\cal T}^{\dagger}}$
and $Y_{\Delta{\cal T}}\equiv Y_{{\cal T}}-Y_{{\cal T}^{\dagger}}$
are \cite{Lavignac:2015gpa}
%%%
\begin{eqnarray}
s{\cal H}z\frac{dY_{\Sigma{\cal T}}}{dz} & = & -\gamma_{D}\left(\frac{Y_{\Sigma{\cal T}}}{Y_{\Sigma{\cal T}}^{{\rm eq}}}-1\right)-2\gamma_{A}\left(\frac{Y_{\Sigma{\cal T}}^{2}}{Y_{\Sigma{\cal T}}^{{\rm eq},2}}-1\right),\\
s{\cal H}z\frac{dY_{\Delta{\cal T}}}{dz} & = & -\gamma_{D}\left(\frac{Y_{\Delta{\cal T}}}{Y_{\Sigma{\cal T}}^{{\rm eq}}}+B_{\ell}\frac{{\rm Tr}\left(ff^{\dagger}Y_{\Delta\ell}\right)}{{\rm Tr}\left(ff^{\dagger}\right)Y^{{\rm nor}}g_{\ell}\zeta_{\ell}}-B_{H}\frac{Y_{\Delta H}}{Y^{{\rm nor}}g_{H}\zeta_{H}}\right),
\end{eqnarray}
%%%
where we have defined $z\equiv M_{{\cal T}}/T$ and to close the equations we apply eqs. (\ref{eq:Yellmatrix_Ycharges})
and (\ref{eq:YH_Ycharges_HiggsTriplets}). The branching ratios for
the decays of ${\cal T}$ to lepton doublets and Higgses are, respectively,
%%%
\begin{eqnarray}
B_{\ell} & = & \frac{{\rm Tr}\left(ff^{\dagger}\right)}{{\rm Tr}\left(ff^{\dagger}\right)+\frac{\left|\mu\right|^{2}}{M_{{\cal T}}^{2}}},\\
B_{H} & = & \frac{\frac{\left|\mu\right|^{2}}{M_{{\cal T}}^{2}}}{{\rm Tr}\left(ff^{\dagger}\right)+\frac{\left|\mu\right|^{2}}{M_{{\cal T}}^{2}}}.
\end{eqnarray}
%%%

For the generation of $Y_{\widetilde{\Delta}}$, we have to append
to the right-hand side of eq. (\ref{eq:BE_YtildeDelta}) a source
and washout terms, respectively, given by \cite{Lavignac:2015gpa}
%%%
\begin{eqnarray}
S^{{\rm II}} & \equiv & -\epsilon\gamma_{D}\left(\frac{Y_{\Sigma{\cal T}}}{Y_{{\cal T}}^{{\rm eq}}}-1\right),\label{eq:source_typeII}\\
W^{{\rm II}} & \equiv & \frac{2\gamma_{D}}{{\rm Tr}\left(ff^{\dagger}\right)+\frac{\left|\mu\right|^{2}}{M_{{\cal T}}^{2}}}\left[\left(ff^{\dagger}\right)\frac{Y_{\Delta{\cal T}}}{Y_{{\cal T}}^{{\rm eq}}}+\frac{1}{4Y^{{\rm nor}}g_{\ell}\zeta_{\ell}}\left(2fY_{\Delta\ell}^{T}f^{\dagger}+ff^{\dagger}Y_{\Delta\ell}+Y_{\Delta\ell}ff^{\dagger}\right)\right],\label{eq:washout_typeII}
\end{eqnarray}
%%%
where the matrix of the $CP$ violation parameter is 
%%%
\begin{eqnarray}
\epsilon & = & \frac{i}{8\pi}\frac{M_{T}}{v^{2}}\sqrt{B_{\ell}B_{H}}\frac{m_{{\cal T}}m_{{\cal H}}^{\dagger}-m_{{\cal H}}m_{{\cal T}}^{\dagger}}{\sqrt{{\rm Tr}\left(m_{{\cal T}}^{\dagger}m_{{\cal T}}\right)}}.\label{eq:CP_typeII}
\end{eqnarray}
%%%
Assuming Maxwell-Boltzmann distribution for ${\cal T}$, we have $Y_{\Sigma{\cal T}}^{{\rm eq}}=Y_{{\cal T}}^{{\rm eq}}+Y_{{\cal T}^{\dagger}}^{{\rm eq}}=\frac{135}{2\pi^{4}g_{\star}}z^{2}{\cal K}_{2}\left(z\right)$
and the decay reaction density $\gamma_{D}$ is given by
%%%
\begin{eqnarray}
\gamma_{D} & = & sY_{\Sigma {\cal T}}^{{\rm eq}}\Gamma_{{\cal T}}\frac{{\cal K}_{1}\left(z\right)}{{\cal K}_{2}\left(z\right)},
\end{eqnarray}
%%%
where the total decay width is
%%%
\begin{eqnarray}
\Gamma_{{\cal T}} & = & \frac{M_{{\cal T}}}{32\pi}\left[{\rm Tr}\left(ff^{\dagger}\right)+\frac{\left|\mu\right|^{2}}{M_{{\cal T}}^{2}}\right].
\end{eqnarray}
%%%

Finally, assuming Maxwell-Boltzmann distributions for all the particles, the gauge scattering reaction density for ${\cal T}{\cal T}^\dagger \leftrightarrow \psi\bar\psi$, where $\psi$ refers to the SM fields, is
%%%
\begin{eqnarray}
\gamma_{A} & = & \frac{M_{{\cal T}}^{4}}{64\pi^{4}z}\int_{4}^{\infty}dx\sqrt{x}{\cal K}_{1}\left(z\sqrt{x}\right)\hat{\sigma}_{A}\left(x\right),
\end{eqnarray}
%%%
where the reduced cross section is given by \cite{Hambye:2005tk}
%%%
\begin{eqnarray}
\hat{\sigma}_{A}\left(x\right) & = & \frac{1}{16\pi x^{2}}\left\{ \sqrt{x}\sqrt{x-4}\left[96g_{2}^{2}g_{Y}^{2}\left(x+4\right)+g_{Y}^{4}\left(65x-68\right)+2g_{2}^{4}\left(172+65x\right)\right]\right.\nonumber \\
 &  & \left.-96\left[4g_{2}^{2}g_{Y}^{2}\left(x-2\right)+g_{Y}^{4}\left(x-2\right)+4g_{2}^{4}\left(x-1\right)\right]\ln\left(\frac{\sqrt{x-4}\sqrt{x}+x}{2}-1\right)\right\} .
\end{eqnarray}
%%%
Taking into account the RGE of the gauge couplings at one loop\footnote{The one-loop RGEs of $\alpha_{2}=\frac{g_{2}^{2}}{4\pi}$
	and $\alpha_{Y}=\frac{g_{Y}^{2}}{4\pi}$ are given by \cite{Gross:1973id,Politzer:1973fx}
	%%%
	\begin{eqnarray*}
	\alpha_{2}\left(\mu\right) & = & \frac{12\pi\alpha_{2}\left(m_{Z}\right)}{12\pi-19\alpha_{2}\left(m_{Z}\right)+19\alpha_{2}\left(m_{Z}\right)\ln\mu},\\
	\alpha_{Y}\left(\mu\right) & = & \frac{20\pi\alpha_{Y}\left(m_{Z}\right)}{20\pi+41\alpha_{Y}\left(m_{Z}\right)-41\alpha_{Y}\left(m_{Z}\right)\ln2\mu}, 
	\end{eqnarray*}
	%%%
	where we take $\mu=2\pi T$ and fix $\alpha_{2}\left(m_{Z}\right)=0.0337$
	and $\alpha_{Y}\left(m_{Z}\right)=0.0169$ with $m_{Z}=91.2$ GeV. }, we obtain an
accurate parametrization within 10\% up to $z\lesssim20$,
%%%
\begin{eqnarray}
\frac{\gamma_{A}}{sHz} & = & \frac{5.5035\times10^{15}\,{\rm GeV}}{g_{\star}^{3/2}M_{{\cal T}}}e^{-1.49z^{1.0735}}.
\end{eqnarray}
%%%

Notice that under flavor rotations in eq. (\ref{eq:flavor_rotations}), from
eqs. (\ref{eq:type_II}) and (\ref{eq:CP_typeII}), we observe that
%%%
\begin{eqnarray}
ff^{\dagger} & \to & V^{*}ff^{\dagger}V^{T},\;\;\;\;\;\epsilon\to V^{*}\epsilon V^{T}.
\end{eqnarray}
%%%
For the source and washout terms \eqref{eq:source_typeII}
and \eqref{eq:washout_typeII} to transform the same way,
%%%
\begin{eqnarray}
S^{{\rm II}} & \to & V^{*}S^{{\rm II}}V^{T},\;\;\;\;\;W^{{\rm II}}\to V^{*}W^{{\rm II}}V^{T},
\end{eqnarray}
%%%
one requires
%%%
\begin{eqnarray}
Y_{\Delta\ell} & \to & V^{*}Y_{\Delta\ell}V^{T}.
\end{eqnarray}
This can be obtained by a particular choice of ordering of flavor
indices as discussed in Appendix \ref{app:matrix_number_densities}.
Hence, we will take $y_{E}\to y_{E}^{*}$ in eqs. (\ref{eq:BE_YE})
and (\ref{eq:BE_YtildeDelta}) such that the transformation is consistent
with the one above. 
Equivalently, we can take $Y_{\Delta \ell} \to Y_{\Delta \ell}^T$ in eqs. (\ref{eq:BE_YE})
and (\ref{eq:BE_YtildeDelta}).

For illustration, we choose a benchmark point from ref. \cite{Lavignac:2015gpa},
%%%
\begin{eqnarray}
m_{{\cal T}} & = & im_{\nu}^{{\rm II}}\implies m_{{\cal H}}=\left(1-i\right)m_{\nu}^{{\rm II}},\\
M_{\Delta} & = & 5\times10^{12}\,{\rm GeV},\\
\left|\mu\right| & = & 0.1M_{{\cal T}},
\end{eqnarray}
%%%
and we fix the neutrino mass matrix to be
%%%
\begin{eqnarray}
m_{\nu} & = & rV_{E}^{T}U_{{\rm PMNS}}^{*}{\rm diag}\left(m_{1},m_{2},m_{3}\right)U_{{\rm PMNS}}^{\dagger}V_{E},
\end{eqnarray}
%%%
with $m_{1}=10^{-3}$ eV while for the rest of the parameters, we choose
the best-fit parameters for normal mass ordering from the global fit
\cite{Esteban:2020cvm}. The effect of RGE up to scale around $M_{\cal T}$ is accounted for approximately by taking $r=1.4$. We ignore the RGE of charge lepton Yukawa and fix it to be
%%%
\begin{equation}
y_{E}=V_{E}^{\dagger}\hat{y}_{E}V_{E},
\end{equation}
%%%
where $\hat{y}_{E}={\rm diag}\left(2.8\times10^{-6},5.9\times10^{-4},1.0\times10^{-2}\right)$.
We will solve the Boltzmann equations in two different bases: nonflavor basis
$y_{E}$ with $V_{E}=U_{{\rm PMNS}}^{\dagger}$ and flavor
basis $\hat{y}_{E}$ with $V_{E}=I_{3\times3}$.

\begin{figure}[H]
	\begin{centering}
		\includegraphics[scale=0.3]{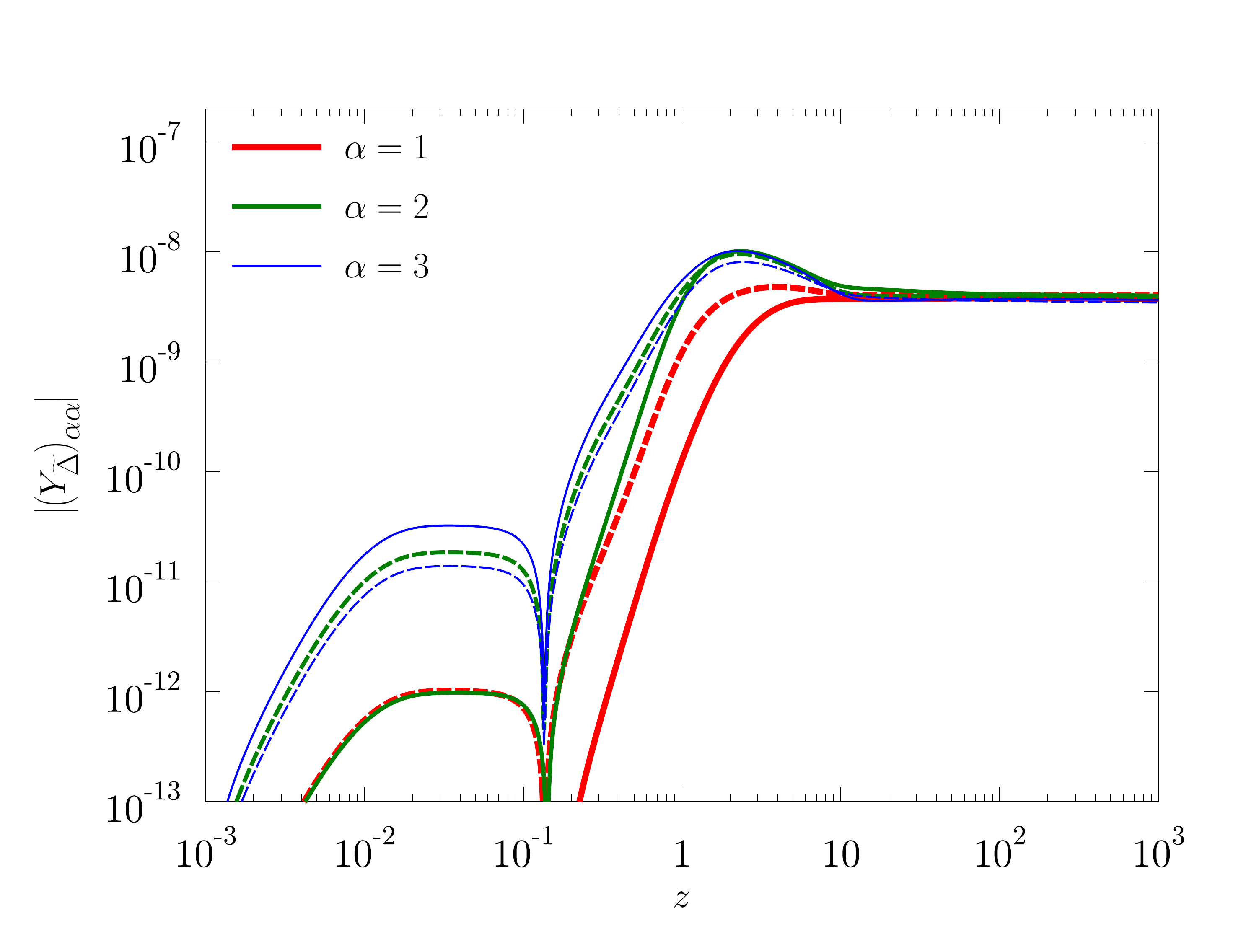}\includegraphics[scale=0.3]{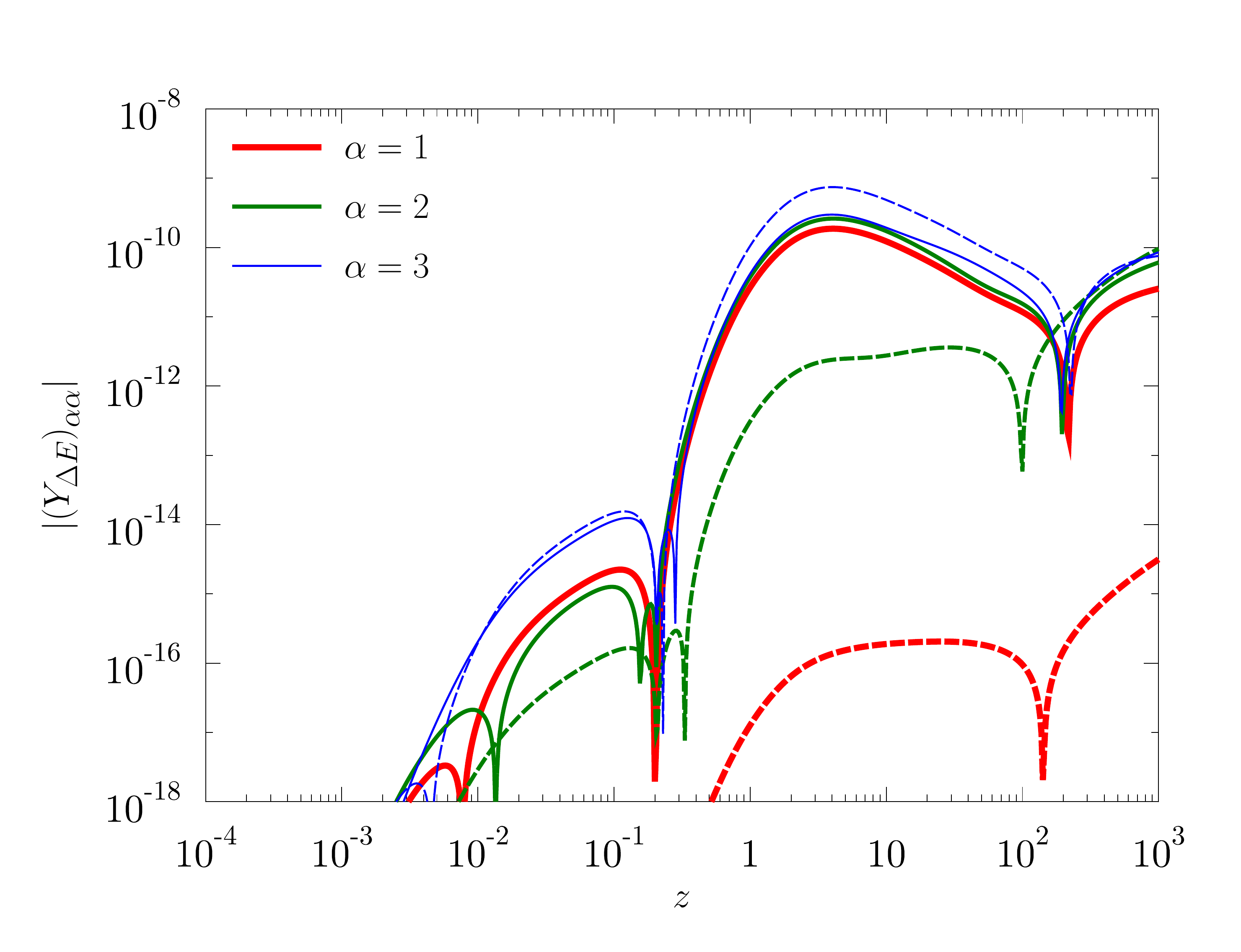}
		\par\end{centering}
	\begin{centering}
		\includegraphics[scale=0.3]{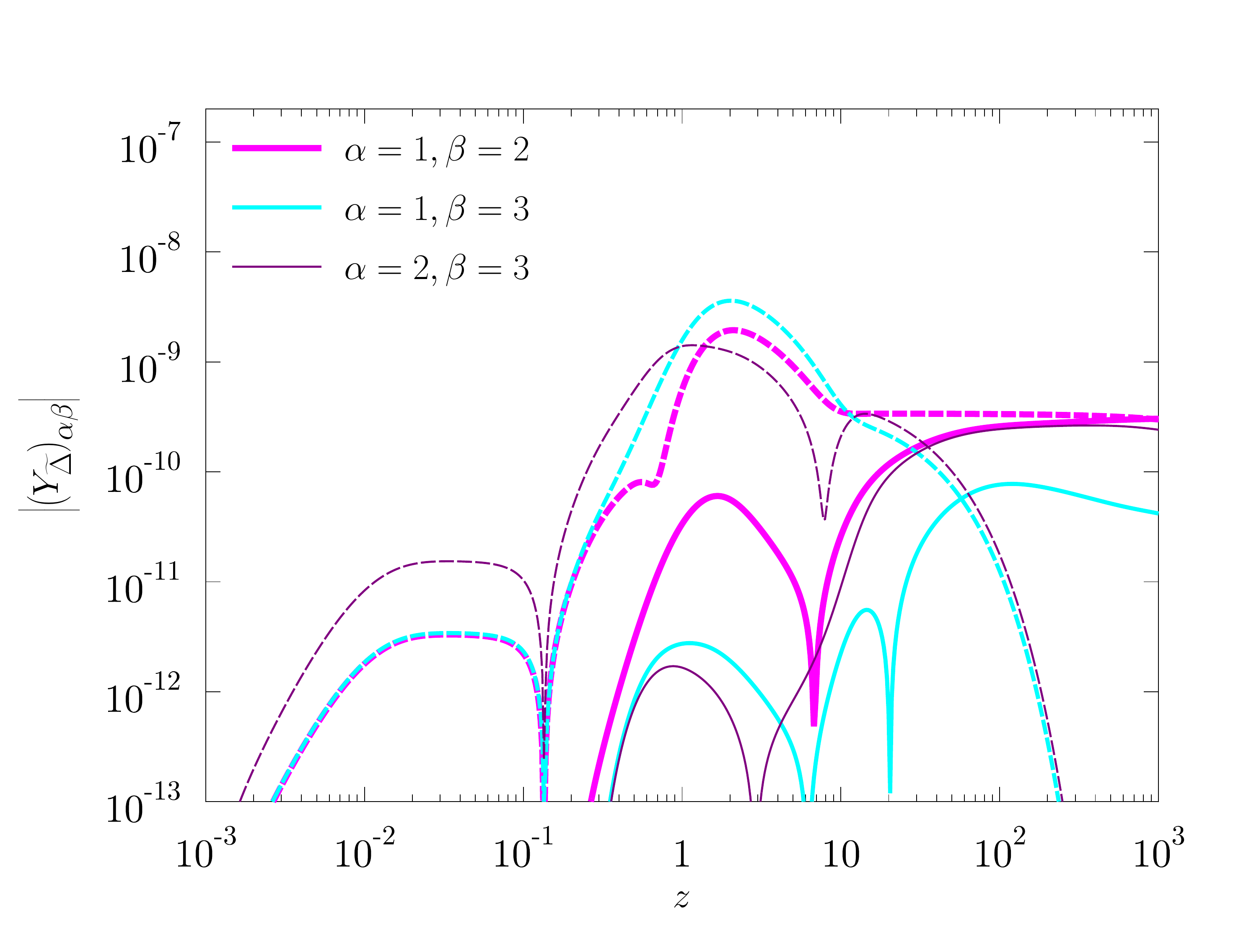}\includegraphics[scale=0.3]{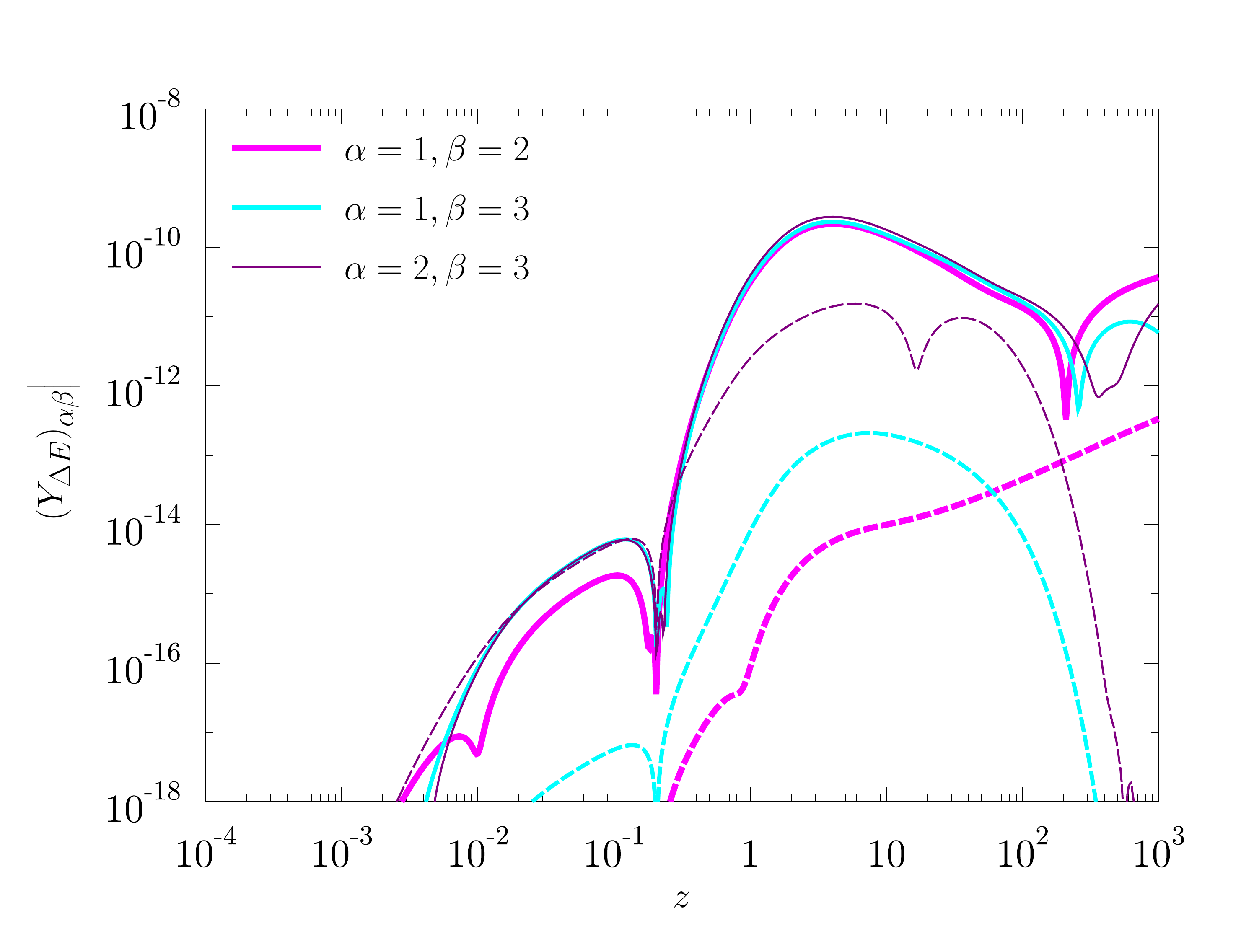}
		\par\end{centering}
	\caption{Numerical solutions for type-II leptogenesis. In the top row, we plot
		the diagonal elements of $\left|Y_{\widetilde{\Delta}}\right|$ and
		$\left|Y_{\Delta E}\right|$ in the two different bases: nonflavor
		basis $y_{E}$ (solid curves) and flavor basis $\hat{y}_{E}$ (dashed
		curves). In the bottom row, we plot the off-diagonal elements of $\left|Y_{\widetilde{\Delta}}\right|$
		and $\left|Y_{\Delta E}\right|$ in the $y_{E}$
		(solid curves) and $\hat{y}_{E}$ (dashed curves) bases. Colors (thickness) denote different matrix elements as indicated in the plots. See the
		text for further discussions. \label{fig:type-II_leptogenesis}}
	
\end{figure}

\begin{figure}[H]
	\begin{centering}
		\includegraphics[scale=0.3]{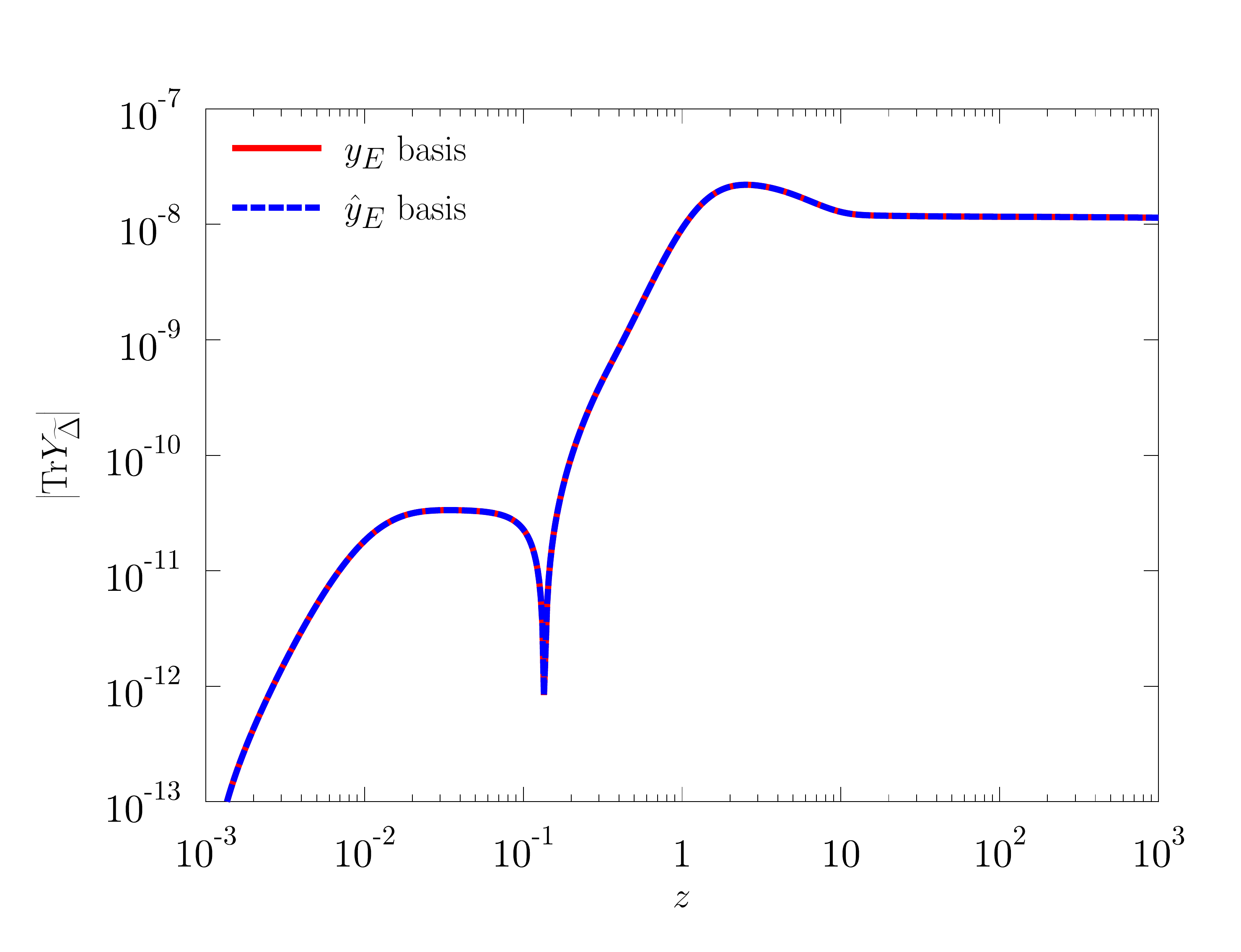}\includegraphics[scale=0.3]{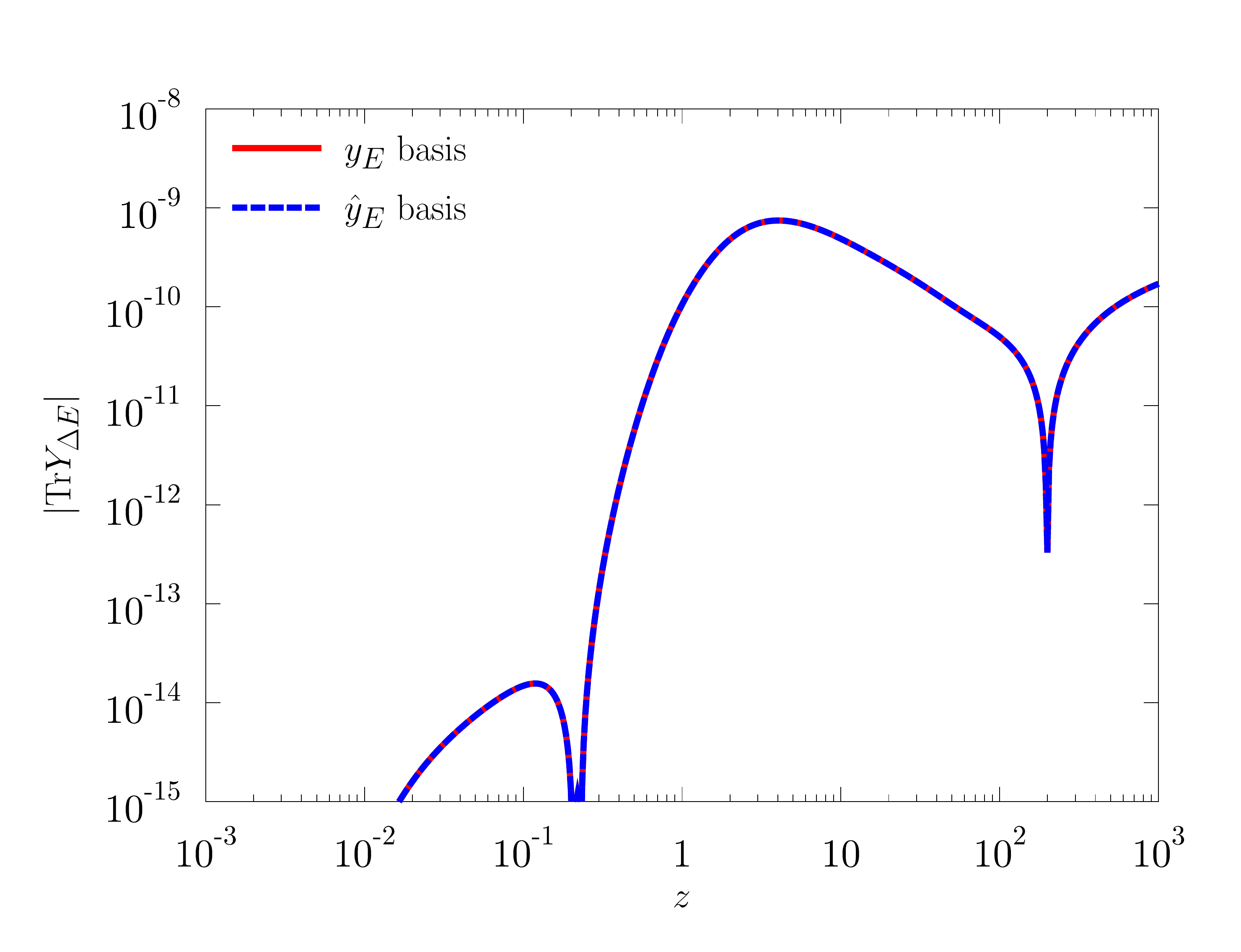}
		\par\end{centering}
	\begin{centering}
		\includegraphics[scale=0.3]{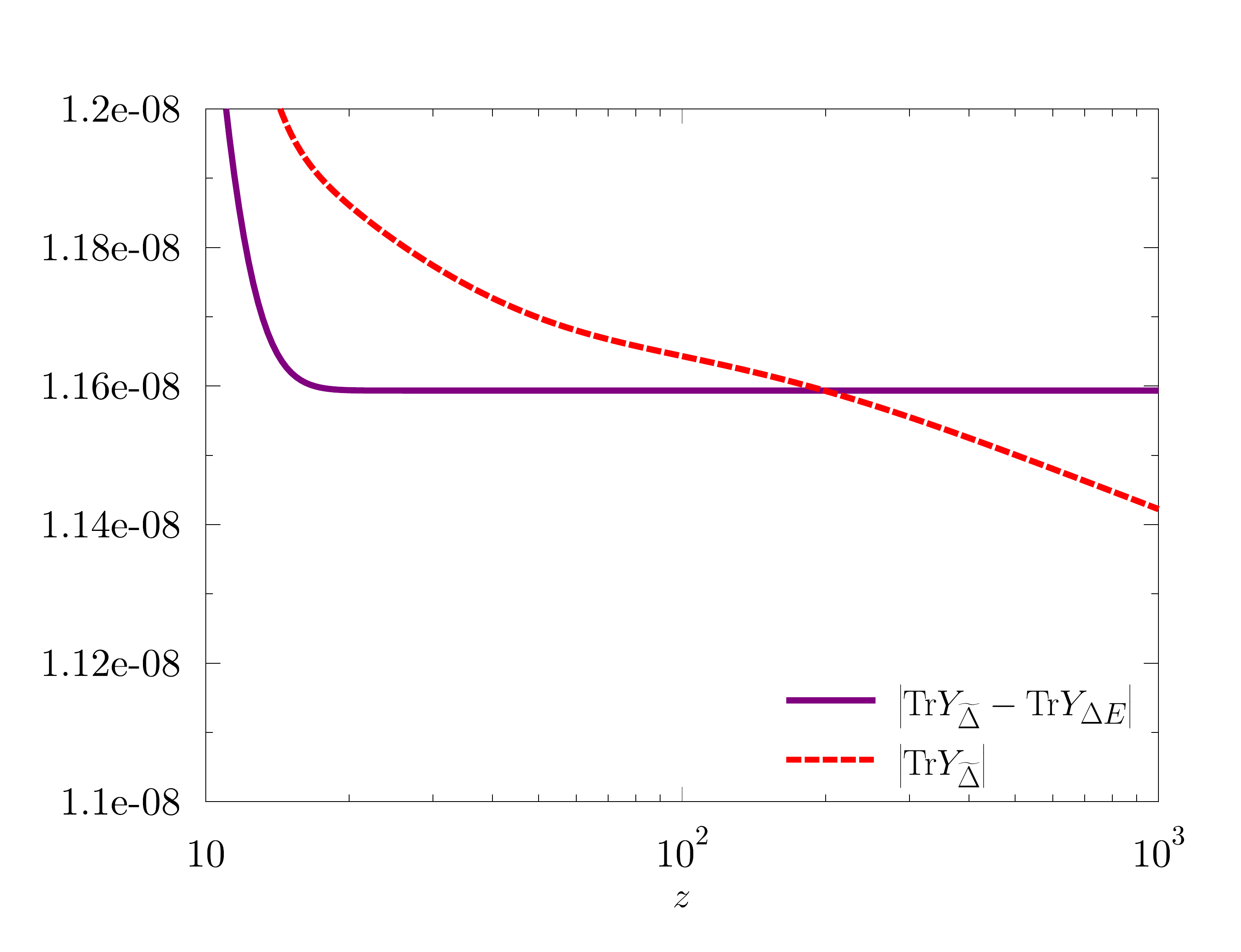}
		\par\end{centering}
	\caption{Numerical solutions for type-II leptogenesis. In the top row, we plot
		$\left|{\rm Tr}Y_{\widetilde{\Delta}}\right|$ and $\left|{\rm Tr}Y_{\Delta E}\right|$
		in the two different bases: nonflavor basis $y_{E}$ (red solid curve)
		and flavor basis $\hat{y}_{E}$ (blue dashed curve). In the bottom
		plot, we plot $\left|{\rm Tr}Y_{\widetilde{\Delta}}-{\rm Tr}Y_{\Delta E}\right|$
		(purple solid curve) and $\left|{\rm Tr}Y_{\widetilde{\Delta}}\right|$
		(red dashed curve) for $z>10$. $\left|{\rm Tr}Y_{\Delta E}\right|$
		is too small to be shown here but one can easily deduced its value from the plot. See the text for further discussions. \label{fig:type-II_leptogenesis-2}}
\end{figure}

In Figure \ref{fig:type-II_leptogenesis}, we show the numerical solutions
comparing the results in $y_{E}$ (solid curves) and $\hat{y}_{E}$ (dashed curves) bases. In the top
row, we show the diagonal elements of $\left|Y_{\widetilde{\Delta}}\right|$
and $\left|Y_{\Delta E}\right|$, while in the bottom row, we show their off-diagonal elements. In the flavor basis $\hat{y}_{E}$,
the off-diagonal elements start to become suppressed at various temperatures
as the charged lepton Yukawa interactions subsequently get into thermal
equilibrium. We see that at $z\sim1000$ $\left(Y_{\widetilde{\Delta}}\right)_{12}$
and $\left(Y_{\Delta E}\right)_{12}$ remain large, indicating that
one has not entered the three-flavor regime. 

In Figure \ref{fig:type-II_leptogenesis-2}, top row, we plot $\left|{\rm Tr}Y_{\widetilde{\Delta}}\right|$
and $\left|{\rm Tr}Y_{\Delta E}\right|$ in the two bases: $y_{E}$
(red solid curve) and $\hat{y}_{E}$ (blue dashed curve). As expected, ${\rm Tr}Y_{\widetilde{\Delta}}$ and ${\rm Tr}Y_{\Delta E}$
are basis independent, while the entries of $Y_{\widetilde{\Delta}}$
and $Y_{\Delta E}$ differ among the two bases by the flavor rotations
as in eq. (\ref{eq:transformation_leptonic_charges}) with $V=U=V_{E}$.
In the bottom plot of Figure \ref{fig:type-II_leptogenesis-2}, we
see that, while $\left|Y_{\Delta(B-L)}\right|=\left|{\rm Tr}Y_{\widetilde{\Delta}}-{\rm Tr}Y_{\Delta E}\right|$
is conserved at the end of leptogenesis, $z\gtrsim20$, while $\left|{\rm Tr}Y_{\widetilde{\Delta}}\right|$
and $\left|{\rm Tr}Y_{\Delta E}\right|$ continue to evolve. In this
example, the final baryon asymmetry obtained is $Y_{\Delta B}\left(T_{B-}\right)=3.6\times10^{-9}$. 

\section{Conclusions\label{sec:Conclusions}}

In this work, we have developed a recipe to describe the evolution of lepton flavor charges from cosmic temperature ranging from $10^{15}$ GeV down to the weak scale, taking into the full SM lepton flavor and spectator effects in a unified and lepton flavor basis-independent manner. This recipe can be applied to any leptogenesis model with the addition of arbitrary number of new scalars with nonzero hypercharges.
We have shown that in order to describe leptogenesis in a basis-independent way and to take into account lepton flavor effect consistently it is necessary to describe both the charges of $\ell$ and $E$ in term of density matrices in their respective flavor spaces. 
To summarize, to apply this formalism to a leptogenesis model is to add the corresponding new physics interactions to the Boltzmann equations \eqref{eq:BE_YE} and \eqref{eq:BE_YtildeDelta}, and then the equations can be closed with eqs. \eqref{eq:Yellmatrix_Ycharges}, \eqref{eq:YH_Ycharges_general}, and those in Appendix \ref{app:quark_relations}.
To demonstrate the applicability of this formalism, we have applied it to type-I and type-II leptogenesis models. Future direction will be to take into account baryon flavor effect.

%%%%%%%%%%%%%%%%%%%%%%%%%%
\section{Acknowledgments}
%%%%%%%%%%%%%%%%%%%%%%%%%%
C.S.F. acknowledges the support by FAPESP Grant No. 2019/11197-6 for the project ``Precision baryogenesis'' and CNPq Grant No. 301271/2019-4. All the Boltzmann equations are solved with Mathematica sponsored by a colleague (who asked not to be named) while all the figures are prepared using Graphics Layout Engine (https://glx.sourceforge.io/). A public code is currently under development.

\appendix

\section{Matrix of number densities\label{app:matrix_number_densities}}

The density matrix operator of the SM in thermal equilibrium at temperature
$T$ is given by
%%%
\begin{eqnarray}
\hat{\rho}_{{\rm SM}} & = & Z^{-1}e^{-\frac{1}{T}\left(\hat{H}_{{\rm SM}}-\sum_{i}\mu_{i}\hat{N}_{i}\right)},
\end{eqnarray}
%%%
where $Z\equiv{\rm Tr}\left[e^{-\frac{1}{T}\left(\hat{H}_{{\rm SM}}-\sum_{i}\mu_{i}\hat{N}_{i}\right)}\right]$
with $\hat{H}_{{\rm SM}}$ the SM Hamiltonian and $\mu_{i}$ and $\hat{N}_{i}$
the chemical potential and number operator of a SM field $i$, respectively.
If we are interested in the correlation between a particle species
of different flavors, we can generalize the chemical potential and
number operators to matrix in flavor space as
%%%
\begin{eqnarray}
\left(\mu_{i}\hat{N}_{i}\right)_{\alpha\beta} & \equiv & 
\int \frac{d^3p}{(2\pi)^3} \left(\mu_{i}\right)_{\alpha\beta}\left(a_{i_{\beta},\mathbf{p}}^{\dagger}a_{i_{\alpha},\mathbf{p}}
-b_{i_{\alpha},\mathbf{p}}^{\dagger}b_{i_{\beta},\mathbf{p}}\right),\;\;\;\;\;\mbox{no sum over $\alpha,\beta$}
\label{eq:number_operator}
\end{eqnarray}
%%%
where $\alpha$ and $\beta$ are flavor indices and we have made use of the
fact that in chemical equilibrium with gauge bosons we have $\mu_{\bar{i}}=-\mu_{i}$.
The operator $a_{i_{\alpha},\mathbf{p}}^{\dagger}$ creates
a particle $i_{\alpha}$ of momentum $\mathbf{p}$ from the vacuum $a_{i_{\alpha},\mathbf{p}}^{\dagger}\left|0\right\rangle =\left|\mathbf{p},i_{\alpha}\right\rangle $
while $b_{i_{\alpha},\mathbf{p}}^{\dagger}$ creates an antilepton
$\bar{i}_{\alpha}$ of momentum $\mathbf{p}$ as $b_{i_{\alpha},\mathbf{p}}^{\dagger}\left|0\right\rangle =\left|\mathbf{p},\bar{i}_{\alpha}\right\rangle $.
For fermions (bosons), they fulfill anticommutator (commutator) relations $\left[ a_{i_{\beta},\mathbf{p}'},a_{i_{\alpha},\mathbf{p}}^{\dagger}\right]_{+(-)} =\left[ b_{i_{\beta},\mathbf{p}'},b_{i_{\alpha},\mathbf{p}}^{\dagger}\right]_{+(-)} =\left(2\pi\right)^{3}\delta^{\left(3\right)}\left(\mathbf{p}-\mathbf{p}'\right)\delta_{\alpha\beta}$
where we have defined $[A,B]_+ \equiv \{A,B\} = AB+BA$ and $[A,B]_- \equiv [A,B] = AB-BA$. Other operator combinations are zero. Sandwiching the operator \eqref{eq:number_operator} between two states with particle of type $i$ of the same momentum $\mathbf{p}$ (but their flavors can be different), we have, for example, $\left<i_\beta,\mathbf{p}\left|\left(\mu_{i}\hat{N}_{i}\right)_{\alpha\beta}\right|i_\alpha,\mathbf{p}\right> = (\mu_i)_{\alpha\beta}$. It also follows that $(\mu_{i})_{\beta\alpha} = (\mu_{i})_{\alpha\beta}^*$.

Next, we will define the generalized phase-space distribution $f_{i(\bar{i})}$
for particle $i$ and antiparticle $\bar{i}$, respectively, as \cite{Sigl:1992fn}
%%%
\begin{eqnarray}
\delta_{\mathbf{p}\mathbf{p}'}\left(f_{i,\mathbf{p}}\right)_{\alpha\beta} & \equiv & {\rm Tr}\left[a_{i_{\beta},\mathbf{p}}^{\dagger}a_{i_{\alpha},\mathbf{p}'}\hat{\rho}_{{\rm SM}}\right],\label{eq:f}\\
\delta_{\mathbf{p}\mathbf{p}'}\left(f_{\bar{i},\mathbf{p}}\right)_{\alpha\beta} & \equiv & {\rm Tr}\left[b_{i_{\alpha},\mathbf{p}}^{\dagger}b_{i_{\beta},\mathbf{p}'}\hat{\rho}_{{\rm SM}}\right],\label{eq:fbar}
\end{eqnarray}
%%%
where we have defined $\delta_{\mathbf{p}\mathbf{p}'} \equiv \left(2\pi\right)^{3}\delta^{\left(3\right)}\left(\mathbf{p}-\mathbf{p}'\right) $.
Notice that the order of flavor indices in $f_{i(\bar{i}),\mathbf{p}}$ determines
how it transforms under flavor rotation of the field. For instance,
considering the fields to be $\ell_{\alpha}$ and $E_{\alpha}$, under
unitary transformations in flavor space, $\ell\to V\ell$ and $E\to UE$,
we have $f_{\ell(\bar{\ell}),\mathbf{p}}\to Vf_{\ell(\bar{\ell}),\mathbf{p}}V^{\dagger}$
and $f_{E(\bar{E}),\mathbf{p}}\to Uf_{E(\bar{E}),\mathbf{p}}U^{\dagger}$. If we have defined
eqs. (\ref{eq:f}) and (\ref{eq:fbar}) with $\left(f_{i,\mathbf{p}}\right)_{\beta\alpha}$
and $\left(f_{\bar{i},\mathbf{p}}\right)_{\beta\alpha}$, they will transform
as $f_{\ell(\bar{\ell}),\mathbf{p}}\to V^{*}f_{\ell(\bar{\ell}),\mathbf{p}}V^{T}$ and $f_{E(\bar{E}),\mathbf{p}}\to U^{*}f_{E(\bar{E}),\mathbf{p}}U^{T}$. Notice that $f_{i,\mathbf{p}}$ and $f_{\bar i,\mathbf{p}}$ are Hermitian.

In what follows, we would like to solve for $\left(f_{i,\mathbf{p}}\right)_{\alpha\beta}$ and
$\left(f_{\bar{i},\mathbf{p}}\right)_{\alpha\beta}$. Since the derivation below follows
independently of whether $\mu_{i}\hat{N}_{i}$ is a matrix in flavor
space or not, we will suppress the flavor indices. Notice that
%%%
\begin{eqnarray}
Z & = & {\rm Tr}\left[e^{-\frac{1}{T}\left(\hat{H}_{{\rm SM}}-\sum_{i}\mu_{i}\hat{N}_{i}\right)}\right]\nonumber \\
 & = & \sum_{{\rm states}}\left\langle {\rm states}\left|e^{-\frac{1}{T}\left(\hat{H}_{{\rm SM}}-\sum_{i}\mu_{i}\hat{N}_{i}\right)}\right|{\rm states}\right\rangle \nonumber \\
 & = & \sum_{{\rm states}}\left\langle {\rm states}\left|e^{-\frac{1}{T}\sum_{i}\left({\cal E}_{i}-\mu_{i}\right)\hat{N}_{i}}\right|{\rm states}\right\rangle \nonumber \\
 & = & \prod_{f}\left[1+e^{-\frac{1}{T}\left({\cal E}_{f}-\mu_{f}\right)}\right]\prod_{b}\left[1-e^{-\frac{1}{T}\left({\cal E}_{b}-\mu_{b}\right)}\right]^{-1}.
\end{eqnarray}
%%%
The traces are taken over multiparticle states with energy ${\cal E}_{i}$.
For fermion $f$, the occupation number is either 0 or 1, and each
of them contributes a factor of $1+e^{-\frac{1}{T}\left({\cal E}_{f}-\mu_{f}\right)}$,
while for boson $b$, each of them contributes a factor of $\sum_{n=0}^{\infty}e^{-\frac{1}{T}\left({\cal E}_{b}-\mu_{b}\right)n}=\left[1-e^{-\frac{1}{T}\left({\cal E}_{b}-\mu_{b}\right)}\right]^{-1}$. 

For a fermion $i$, we have
%%%
\begin{eqnarray}
{\rm Tr}\left[a_{i,\mathbf{p}}^{\dagger} a_{i,\mathbf{p}'}\hat{\rho}_{{\rm SM}}\right] & = & 
Z^{-1}\delta_{\mathbf{p}\mathbf{p}'}
e^{-\frac{1}{T}\left({\cal E}_{i}-\mu_{i}\right)}\nonumber \\
 &  & \times\prod_{f\neq i}\left[1+e^{-\frac{1}{T}\left({\cal E}_{f}-\mu_{f}\right)}\right]\prod_{b}\left[1-e^{-\frac{1}{T}\left({\cal E}_{b}-\mu_{b}\right)}\right]^{-1}\nonumber \\
 & = & \delta_{\mathbf{p}\mathbf{p}'}\frac{e^{-\frac{1}{T}\left({\cal E}_{i}-\mu_{i}\right)}}{1+e^{-\frac{1}{T}\left({\cal E}_{i}-\mu_{i}\right)}}\nonumber \\
 & = & \delta_{\mathbf{p}\mathbf{p}'}\frac{1}{e^{\frac{1}{T}\left({\cal E}_{i}-\mu_{i}\right)}+1}.
\end{eqnarray}
%%%

For a boson $i$, we have
%%%
\begin{eqnarray}
{\rm Tr}\left[a_{i,\mathbf{p}}^{\dagger}a_{i,\mathbf{p}'}\hat{\rho}_{{\rm SM}}\right] & = & Z^{-1}
\delta_{\mathbf{p}\mathbf{p}'}
%\left(2\pi\right)^{3}\delta^{\left(3\right)}\left(\vec{p}-\vec{p}'\right)
\sum_{n}ne^{-\frac{1}{T}\left({\cal E}_{i}-\mu_{i}\right)n}\nonumber \\
 &  & \times\prod_{f}\left[1+e^{-\frac{1}{T}\left({\cal E}_{f}-\mu_{f}\right)}\right]\prod_{b\neq i}\left[1-e^{-\frac{1}{T}\left({\cal E}_{b}-\mu_{b}\right)}\right]^{-1}\nonumber \\
 & = & Z^{-1}\delta_{\mathbf{p}\mathbf{p}'}
 \frac{e^{-\frac{1}{T}\left({\cal E}_{i}-\mu_{i}\right)}}{\left[1-e^{-\frac{1}{T}\left({\cal E}_{i}-\mu_{i}\right)}\right]^{2}}\nonumber \\
 &  & \times\prod_{f}\left[1+e^{-\frac{1}{T}\left({\cal E}_{f}-\mu_{f}\right)}\right]\prod_{b\neq i}\left[1-e^{-\frac{1}{T}\left({\cal E}_{b}-\mu_{b}\right)}\right]^{-1}\nonumber \\
 & = & \delta_{\mathbf{p}\mathbf{p}'}
 \frac{e^{-\frac{1}{T}\left({\cal E}_{i}-\mu_{i}\right)}}{\left[1-e^{-\frac{1}{T}\left({\cal E}_{i}-\mu_{i}\right)}\right]^{2}}\left[1-e^{-\frac{1}{T}\left({\cal E}_{i}-\mu_{i}\right)}\right]\nonumber \\
 & = & \delta_{\mathbf{p}\mathbf{p}'}
 \frac{1}{e^{\frac{1}{T}\left({\cal E}_{i}-\mu_{i}\right)}-1}.
\end{eqnarray}
%%%
In the second line above, we have used $\sum_{n}ne^{-\frac{1}{T}\left({\cal E}_{i}-\mu_{i}\right)n}
=-\frac{1}{\frac{1}{T}\left({\cal E}_{i}-\mu_{i}\right)}\frac{d}{dn}\sum_{n}e^{-\frac{1}{T}\left({\cal E}_{i}-\mu_{i}\right)n}
=\frac{e^{-\frac{1}{T}\left({\cal E}_{i}-\mu_{i}\right)}}{\left[1-e^{-\frac{1}{T}\left({\cal E}_{i}-\mu_{i}\right)}\right]^{2}}$. 

One can repeat the exercise above for antiparticle $\bar i$ with $a_{i}\to b_{i}$ and the only change is $\mu_{i}\to-\mu_{i}$. Hence, from the definitions (\ref{eq:f})
and (\ref{eq:fbar}), we obtain the desired results
%%%
\begin{eqnarray}
\left(f_{i,\mathbf{p}}\right)_{\alpha\beta} & = & \frac{1}{e^{\frac{{\cal E}_{i}-\left(\mu_{i}\right)_{\alpha\beta}}{T}}+\xi_{i}},
\;\;\;\;\;
\left(f_{\bar{i},\mathbf{p}}\right)_{\alpha\beta} = \frac{1}{e^{\frac{{\cal E}_{i}+\left(\mu_{i}\right)_{\alpha\beta}}{T}}+\xi_{i}},
\label{eq:f_i_ibar}
\end{eqnarray}
%%%
where $\xi_{i}=1(-1)$ for $i$ a fermion (boson) and ${\cal E}_i = \sqrt{|\mathbf{p}|^2+m_i^2}$. Integrating the phase space distributions above over 3-momentum, we obtain (matrices) of number densities
%%%
\begin{eqnarray}
\left(n_{i}\right)_{\alpha\beta} & \equiv &
g_i \int \frac{d^3 p}{(2\pi)^3} 
\left(f_{i,\mathbf{p}}\right)_{\alpha\beta},
\;\;\;\;\;
\left(n_{\bar i}\right)_{\alpha\beta} \equiv 
g_i \int \frac{d^3 p}{(2\pi)^3}  
\left(f_{\bar i,\mathbf{p}}\right)_{\alpha\beta}.
\end{eqnarray}
%%%
where we have included $g_{i}$ to take into account additional gauge degrees of freedom.

Expanding to linear order in chemical potential $\left|\mu_{i}\right|/T\ll1$ and integrating over 3-momentum, the difference between the phase-space distributions of $i$ and $\bar i$, we obtain the (matrix of the) number density asymmetry
%%%
\begin{eqnarray}
\left(n_{\Delta i}\right)_{\alpha\beta} & \equiv & 
g_i \int \frac{d^3 p}{(2\pi)^3} 
\left[\left(f_{i,\mathbf{p}}\right)_{\alpha\beta} 
- \left(f_{\bar i,\mathbf{p}}\right)_{\alpha\beta}\right]
%n_{i}-n_{\bar{i}}
=\frac{T^{2}}{6}g_{i}\zeta_{i}\left(\mu_{i}\right)_{\alpha\beta},
\label{eq:density_asymmetry_chempot}
\end{eqnarray}
%%%
where we have defined
%%%
\begin{eqnarray}
\zeta_{i} & \equiv & \frac{6}{\pi^{2}}\int_{m_{i}/T}^{\infty}dx\,x\sqrt{x^{2}-m_{i}^{2}/T^{2}}\frac{e^{x}}{\left(e^{x}+\xi_{i}\right)^{2}}.
\end{eqnarray}
%%%
For massless particle $m_{i}=0$, we have $\zeta_{i}=1(2)$ for $i$
a massless fermion (boson). 
The transformation of $\left(n_{\Delta i}\right)_{\alpha\beta}$ follows
directly from eqs. (\ref{eq:f}) and (\ref{eq:fbar}). For instance,
considering the fields to be $\ell_{\alpha}$ and $E_{\alpha}$, under
unitary transformations in flavor space, $\ell\to V\ell$ and $E\to UE$,
we have $n_{\Delta\ell}\to Vn_{\Delta\ell}V^{\dagger}$ and $n_{\Delta E}\to Un{}_{\Delta E}U^{\dagger}$.
Alternatively, if we have defined eqs. (\ref{eq:f}) and (\ref{eq:fbar})
with $\left(f_{i,\mathbf{p}}\right)_{\beta\alpha}$ and $\left(f_{\bar{i},\mathbf{p}}\right)_{\beta\alpha}$, the transformations will be $n_{\Delta\ell}\to V^{*}n_{\Delta\ell}V^{T}$
and $n_{\Delta E}\to U^{*}n_{\Delta E}U^{T}$.

Normalizing eq. \eqref{eq:density_asymmetry_chempot} by the cosmic entropy density $s=\frac{2\pi^{2}}{45}g_{\star}T{{}^3}$
with $g_{\star}$ being the effective relativistic degrees of freedom of the Universe, we have
%%%
\begin{eqnarray}
\left(Y_{\Delta i}\right)_{\alpha\beta} & \equiv & \frac{\left(n_{\Delta i}\right)_{\alpha\beta}}{s}\equiv Y^{{\rm nor}}g_{i}\zeta_{i}\frac{2(\mu_{i})_{\alpha\beta}}{T},
\label{eq:Y_mu}
\end{eqnarray}
%%%
where we have defined $Y^{\rm nor} \equiv \frac{15}{8\pi^2 g_\star}$. The relation above also holds for a particle which does not carry a flavor index, e.g., for the SM Higgs, which is taken to be massless at high temperature, we have  $Y_{\Delta H} = 4 Y^{\rm nor} \frac{2\mu_H}{T}$, where $g_H = 2$ for the $SU(2)_L$ gauge degrees of freedom and $\xi_H = 2$ for massless boson.

\section{Covariant flavor structures of kinetic equations \label{app:flavor_kinetic_equations}}

The complete flavor-covariant kinetic equations have been derived in refs. \cite{Beneke:2010dz,Garbrecht:2013bia} using the closed time path formalism. Here we would like to sketch how the same flavor structures of the kinetic equations for $\ell$ and $E$ arise by considering the evolution equation of a Heisenberg operator \cite{Sigl:1992fn}.

We will start by deriving some relations relating the equilibrium phase-space distributions
with the (matrices) of number density asymmetries. As shown the previous
section, for a particle $i$ which is in kinetic equilibrium, its
phase-space distribution is given by eq. \eqref{eq:f_i_ibar}. For a process $ab...\leftrightarrow ij...$
which is in chemical equilibrium $\mu_{a}+\mu_{b}+...=\mu_{i}+\mu_{j}+...$
[if a particle carries a family index, e.g., $i_{\alpha}$, the chemical
potential refers to the corresponding diagonal element $\left(\mu_{i}\right)_{\alpha\alpha}$],
we can verify that the following identity is satisfied,
%%%
\begin{eqnarray}
f_{a}f_{b}...\left(1-\xi_{i}f_{i}\right)\left(1-\xi_{j}f_{j}\right)... & = & f_{i}f_{j}...\left(1-\xi_{a}f_{a}\right)\left(1-\xi_{b}f_{b}\right)...,\label{eq:identity}
\end{eqnarray}
%%%
where we have used energy conservation ${\cal E}_{a}+{\cal E}_{b}+...={\cal E}_{i}+{\cal E}_{j}+...$
and we have suppressed the subscript of momentum/energy in the distribution
functions. In general, we cannot make use of the identity above since
chemical equilibrium condition is not necessarily fulfilled when the
corresponding process is slower than the Hubble expansion rate. Defining the distribution of a particle $i$ in kinetic equilibrium with zero chemical potential as $f_{i}^{{\rm eq}}$, the following identity is clearly satisfied 
%%%
\begin{eqnarray}
f_{a}^{{\rm eq}}f_{b}^{{\rm eq}}...\left(1-\xi_{i}f_{i}^{{\rm eq}}\right)\left(1-\xi_{j}f_{j}^{{\rm eq}}\right)... & = & f_{i}^{{\rm eq}}f_{j}^{{\rm eq}}...\left(1-\xi_{a}f_{a}^{{\rm eq}}\right)\left(1-\xi_{b}f_{b}^{{\rm eq}}\right)....\label{eq:identity_eq}
\end{eqnarray}
%%%

In the following, let us consider all the particles are in kinetic equilibrium (this holds for all the SM particles which experience gauge interactions). Expanding in chemical potentials $\left|\mu_{i}\right|/T\ll1$
up to linear order for all the particles, we have
%%%
\begin{eqnarray}
\frac{f_{a}f_{b}...\left(1-\xi_{i}f_{i}\right)\left(1-\xi_{j}f_{j}\right)...}{f_{a}^{{\rm eq}}f_{b}^{{\rm eq}}...\left(1-\xi_{i}f_{i}^{{\rm eq}}\right)\left(1-\xi_{j}f_{j}^{{\rm eq}}\right)...} & = & \left[1+\frac{\mu_{a}}{T}\left(1-\xi_{a}f_{a}^{{\rm eq}}\right)+\frac{\mu_{b}}{T}\left(1-\xi_{b}f_{b}^{{\rm eq}}\right)+...\right]\nonumber \\
&  & \times\left(1-\frac{\mu_{i}}{T}\xi_{i}f_{i}^{{\rm eq}}-\frac{\mu_{j}}{T}\xi_{j}f_{j}^{{\rm eq}}+...\right)\nonumber \\
& = & 1+\sum_{I=a,b,...}\frac{\mu_{I}}{T}-\sum_{I=a,b,...}\frac{\mu_{I}}{T}\xi_{I}f_{I}^{{\rm eq}}-\sum_{F=i,j,...}\frac{\mu_{F}}{T}\xi_{F}f_{F}^{{\rm eq}}\nonumber \\
& = & 1+\sum_{I=a,b,...}\frac{Y_{\Delta I}}{2Y^{{\rm nor}}g_{I}\zeta_{I}}-\sum_{A=a,b,...,i,j,...}\frac{Y_{\Delta A}\xi_{A}f_{A}^{{\rm eq}}}{2Y^{{\rm nor}}g_{A}\zeta_{A}},\label{eq:expansion1}
\end{eqnarray}
%%%
where in the last step we have used eq. \eqref{eq:Y_mu}. Similarly, we have the relation
for antiparticles by changing the sign of the chemical potentials,
%%%
\begin{eqnarray}
\frac{f_{\bar{a}}f_{\bar{b}}...\left(1-\xi_{i}f_{\bar{i}}\right)\left(1-\xi_{j}f_{\bar{j}}\right)...}{f_{a}^{{\rm eq}}f_{b}^{{\rm eq}}...\left(1-\xi_{i}f_{i}^{{\rm eq}}\right)\left(1-\xi_{j}f_{j}^{{\rm eq}}\right)...} & = & 1-\sum_{I=a,b,...}\frac{Y_{\Delta I}}{2Y^{{\rm nor}}g_{I}\zeta_{I}}+\sum_{A=a,b,...,i,j,...}\frac{Y_{\Delta A}\xi_{A}f_{A}^{{\rm eq}}}{2Y^{{\rm nor}}g_{A}\zeta_{A}}.\label{eq:expansion2}
\end{eqnarray}
%%%

The evolution equations of the Heisenberg operators $\left({\cal O}_{i,\mathbf{p}}\right)_{\alpha\beta}(t)\equiv a_{i_{\beta},\mathbf{p}}^{\dagger}(t)a_{i_{\alpha},\mathbf{p}}(t)$
and $\left({\cal O}_{\bar{i},\mathbf{p}}\right)_{\alpha\beta}(t)\equiv b_{i_{\alpha},\mathbf{p}}^{\dagger}(t)b_{i_{\beta},\mathbf{p}}(t)$
are given by
%%%
\begin{eqnarray}
\frac{\partial\left({\cal O}_{i,\mathbf{p}}\right)_{\alpha\beta}}{\partial t} & = & i\left[\hat H,\left({\cal O}_{i,\mathbf{p}}\right)_{\alpha\beta}\right],\label{eq:EOM_Oi}\;\;\;\;\;
\frac{\partial\left({\cal O}_{\bar{i},\mathbf{p}}\right)_{\alpha\beta}}{\partial t} = i\left[\hat H,\left({\cal O}_{\bar{i},\mathbf{p}}\right)_{\alpha\beta}\right],\label{eq:EOM_Oibar}
\end{eqnarray}
%%%
where $\hat H= \hat H_{0}+\hat H_{{\rm int}}$ is the Hamiltonian of the system with
$H_{0}$ denoting the free field Hamiltonian while $\hat H_{{\rm int}}$
represents all possible interactions among the fields. In the following,
we will write down the derivation only for the equation of motion
of ${\cal O}_{i,\mathbf{p}}$ since those for ${\cal O}_{\bar{i},\mathbf{p}}$
will be analogous.

Taking the ensemble average on both sides of eq. (\ref{eq:EOM_Oi}),
we have 
%%%
\begin{eqnarray}
\frac{\partial\left(f_{i,\mathbf{p}}\right)_{\alpha\beta}}{\partial t} & = & i\left\langle \left[\hat H,\left({\cal O}_{i,\mathbf{p}}\right)_{\alpha\beta}\right]\right\rangle ,
\end{eqnarray}
%%%
where we have denoted $\left\langle {\cal O}\right\rangle \equiv{\rm Tr}\left[{\cal O}\hat{\rho}_{{\rm SM}}\right]$.
The effect of cosmic expansion can be taken into account by adding the following
%%%
\begin{eqnarray}
\frac{\partial\left(f_{i,\mathbf{p}}\right)_{\alpha\beta}}{\partial t}-{\cal H}\left|\mathbf{p}\right|\frac{\partial\left(f_{i,\mathbf{p}}\right)_{\alpha\beta}}{\partial\left|\mathbf{p}\right|} & = & i\left\langle \left[\hat H,\left({\cal O}_{i,\mathbf{p}}\right)_{\alpha\beta}\right]\right\rangle .
\end{eqnarray}
%%%
Integrating the equation above over momentum $\mathbf{p}$ on both
sides, we have
%%%
\begin{eqnarray}
\frac{d\left(n_{i}\right)_{\alpha\beta}}{dt}+3{\cal H}\left(n_{i}\right)_{\alpha\beta} & = & i\int\frac{d^{3}p}{\left(2\pi\right)^{3}}\left\langle \left[\hat H,\left({\cal O}_{i,\mathbf{p}}\right)_{\alpha\beta}\right]\right\rangle,
\label{eq:Heisenberg_equation}
\end{eqnarray}
%%%
where we have defined the number density (matrix) as
%%%
\begin{eqnarray}
\left(n_{i}\right)_{\alpha\beta} & \equiv & \int\frac{d^{3}p}{\left(2\pi\right)^{3}}\left(f_{i,\mathbf{p}}\right)_{\alpha\beta},
\end{eqnarray}
%%%
and assume that $f_{i,\mathbf{p}}$ goes to zero at large momentum. In the absence of interactions $\hat H_{{\rm int}}=0$, the phase space will evolve purely due to the Hubble expansion. In terms of $Y_{i}\equiv n_{i}/s$,
we can rewrite
%%%
\begin{eqnarray}
\frac{d\left(n_{i}\right)_{\alpha\beta}}{dt}+3{\cal H}\left(n_{i}\right)_{\alpha\beta} & = & s\frac{d\left(Y_{i}\right)_{\alpha\beta}}{dt}.
\end{eqnarray}
%%%

For massless fields, $H_{0}$ does not contribution to the right-hand side of eq. \eqref{eq:Heisenberg_equation}.
Next, we would like to write the terms in right-hand side of evolution
equation also in terms of number densities. Doing a perturbative expansion
on the Heisenberg operator $\left[\hat H_{\rm int},\left({\cal O}_{i,\mathbf{p}}\right)_{\alpha\beta}\right]$
to the first order in $\hat H_{{\rm int}}$, and considering that the interaction
timescale is much shorter than the evolution timescale, we can take
the time integral to infinity and obtain \cite{Sigl:1992fn}
%%%
\begin{eqnarray}
s\frac{d\left(Y_{i}\right)_{\alpha\beta}}{dt} & = & i\int\frac{d^{3}p}{\left(2\pi\right)^{3}}\left\langle \left[\hat H_{{\rm int},0}\left(0\right),\left({\cal O}_{i,\mathbf{p}}\right)_{\alpha\beta,0}\left(0\right)\right]\right\rangle \nonumber \\
&  & -\int\frac{d^{3}p}{\left(2\pi\right)^{3}}\int_{0}^{\infty}dt\left\langle \left[\hat H_{{\rm int},0}\left(t\right),\left[\hat H_{{\rm int},0}\left(0\right),\left({\cal O}_{i,\mathbf{p}}\right)_{\alpha\beta,0}\left(0\right)\right]\right]\right\rangle ,\label{eq:evolution_expansion}
\end{eqnarray}
%%%
where the subscript 0 denote operators consist of free fields i.e.
$\hat H_{{\rm int}}=0$. 

Considering only the SM charged lepton Yukawa interaction term, we have 
%%%
\begin{eqnarray}
\hat H_{{\rm int}} & = & \int d^{3}x\left[\left(y_{E}\right)_{\alpha\beta}\overline{E_{\alpha}}\ell_{\beta}H^{*}+\left(y_{E}\right)_{\alpha\beta}^{*}\overline{\ell_{\beta}}E_{\alpha}H\right],
\end{eqnarray}
%%%
Since this interaction is linear in the each type of field, it will only contribute to the second term. Considering thermal mass \cite{Weldon:1982bn}, there is a contribution to the first term of eq. \eqref{eq:evolution_expansion}, which results in oscillation among $\ell$ flavors. Ref. \cite{Beneke:2010dz} showed that flavor oscillations are damped by gauge interactions, and hence we will ignore this term.

Expanding the fields in momentum modes, we have
%%%
\begin{eqnarray}
\ell_{\beta} & = & \int\frac{d^{3}p}{\left(2\pi\right)^{3}\sqrt{2{\cal E}}}\sum_{s}\left(a_{\ell_{\beta},\mathbf{p}}^{(s)}u_{\ell,\mathbf{p}}^{(s)}e^{-ip\cdot x}+b_{\ell_{\beta},\mathbf{p}}^{(s)\dagger}v_{\ell,\mathbf{p}}^{(s)}e^{+ip\cdot x}\right),\\
E_{\alpha} & = & \int\frac{d^{3}p}{\left(2\pi\right)^{3}\sqrt{2{\cal E}}}\sum_{s}\left(a_{E_{\alpha},\mathbf{p}}^{(s)}u_{E,\mathbf{p}}^{(s)}e^{-ip\cdot x}+b_{E_{\alpha},\mathbf{p}}^{(s)\dagger}v_{E,\mathbf{p}}^{(s)}e^{+ip\cdot x}\right),\\
H & = & \int\frac{d^{3}p}{\left(2\pi\right)^{3}\sqrt{2{\cal E}}}\left(a_{H,\mathbf{p}}e^{-ip\cdot x}+b_{H,\mathbf{p}}^{\dagger}e^{+ip\cdot x}\right),
\end{eqnarray}
%%%
where $p\cdot x={\cal E}t-\mathbf{p}\cdot\mathbf{x}$ and the sum
$s$ is taken over the two spin states. Substituting the fields
above into eq. (\ref{eq:evolution_expansion}), we obtain the evolution
equation of $\ell$ as
%%%
\begin{eqnarray}
s\frac{d\left(Y_{\ell}\right)_{\alpha\beta}}{dt} & = & -\frac{1}{2}\int\frac{d^{3}p}{\left(2\pi\right)^{3}2{\cal E}_{\ell}}\frac{d^{3}p_{E}}{\left(2\pi\right)^{3}2{\cal E}_{E}}\frac{d^{3}p_{H}}{\left(2\pi\right)^{3}2{\cal E}_{H}}2p\cdot p_{E} \label{eq:evolution_nell}\\
&  &\hspace{-1.1cm} \left\{ \delta_{p_{E}-p_{\ell}+p_{H}}\left[\left\{ f_{\ell,\mathbf{p}},y_{E}^{\dagger}\left(I-f_{E,\mathbf{p}_{E}}\right)y_{E}\right\} _{\alpha\beta}\left(1+f_{H,\mathbf{p}_{H}}\right)-\left\{ \left(I-f_{\ell,\mathbf{p}}\right),y_{E}^{\dagger}f_{E,\mathbf{p}_{E}}y_{E}\right\} _{\alpha\beta}f_{H,\mathbf{p}_{H}}\right]\right.\nonumber \\
&  &\hspace{-1.1cm} +\delta_{p_{E}-p_{\ell}-p_{H}}\left[\left\{ f_{\ell,\mathbf{p}},y_{E}^{\dagger}\left(I-f_{E,\mathbf{p}_{E}}\right)y_{E}\right\} _{\alpha\beta}f_{\overline{H},\mathbf{p}_{H}}-\left\{ \left(I-f_{\ell,\mathbf{p}}\right),y_{E}^{\dagger}f_{E,\mathbf{p}_{E}}y_{E}\right\} _{\alpha\beta}\left(1+f_{\overline{H},\mathbf{p}_{H}}\right)\right]\nonumber \\
&  &\hspace{-1.1cm} +\delta_{p_{E}+p_{\ell}-p_{H}}\left[\left\{ f_{\ell,\mathbf{p}},y_{E}^{\dagger}f_{\overline{E},\mathbf{p}_{E}}y_{E}\right\} _{\alpha\beta}\left(1+f_{H,\mathbf{p}_{H}}\right)-\left\{ \left(I-f_{\ell,\mathbf{p}}\right),y_{E}^{\dagger}\left(I-f_{\overline{E},\mathbf{p}_{E}}\right)y_{E}\right\} _{\alpha\beta}f_{H,\mathbf{p}_{H}}\right]\nonumber \\
&  &\hspace{-1.1cm} \left.+\delta_{p_{E}+p_{\ell}+p_{H}}\left[\left\{ f_{\ell,\mathbf{p}},y_{E}^{\dagger}f_{\overline{E},\mathbf{p}_{E}}y_{E}\right\} _{\alpha\beta}f_{\overline{H},\mathbf{p}_{H}}-\left\{ \left(I-f_{\ell,\mathbf{p}}\right),y_{E}^{\dagger}\left(I-f_{\overline{E},\mathbf{p}_{E}}\right)y_{E}\right\} _{\alpha\beta}\left(1+f_{\overline{H},\mathbf{p}_{H}}\right)\right]\right\} ,\nonumber
\end{eqnarray}
%%%
where $\delta_{p}\equiv\left(2\pi\right)^{4}\delta^{(4)}(p)$ with
$p$ a 4-momentum, $I$ is a $3\times3$ identity matrix and we
have assumed all external fields to be massless. Clearly, the whole
term vanishes since $p\cdot p_{E}\propto p_{H}^{2}=0$. 

For \emph{nonvanishing} result, one should consider thermal masses and scattering processes involving another external field \cite{Cline:1993bd}. For instance, a gauge field can be attached to either $\ell$, $E$
or $H$. One can also attach a fermion-antifermion
pair to the Higgs fields and the process involving the top-quark Yukawa coupling will be
the dominant one. Since the flavor structures involving $\ell$ and $E$ will remain exactly the same, we will not carry out the exercise here.\footnote{For the scatterings involving two quarks instead of the Higgs fields, one can still rewrite in terms of Higgs number density asymmetry for $T < T_t \sim 10^{15}$ GeV when interactions involving the top-quark Yukawa are in equilibrium.}
Notice that the last term in the big curly brackets of eq. \eqref{eq:evolution_nell} will still be zero due to energy-momentum conservation.

Dividing and multiplying the terms in the right-hand side of eq. \eqref{eq:evolution_nell} by $f_{\ell,\mathbf{p}}^{{\rm eq}}\left(1-f_{E,\mathbf{p}_{E}}^{{\rm eq}}\right)\left(1+f_{H,\mathbf{p}_{H}}^{{\rm eq}}\right)=f_{E,\mathbf{p}_{E}}^{{\rm eq}}f_{H,\mathbf{p}_{H}}^{{\rm eq}}\left(1-f_{\ell,\mathbf{p}}^{{\rm eq}}\right)$ which follows from eq. \eqref{eq:identity_eq} and expanding up to
linear term in $\left|\mu_{i}\right|/T\ll1$, the first three terms
in the big curly brackets in eq. (\ref{eq:evolution_nell}) all have
the same flavor structure,
%%%
\begin{eqnarray}
s\frac{dY_{\ell}}{dt} & \sim & -\frac{1}{2Y^{{\rm nor}}}\left[\left\{ y_{E}^{\dagger}y_{E},\frac{Y_{\Delta\ell}}{g_{\ell}\zeta_{\ell}}\right\}-2\left(\frac{y_{E}^{\dagger}Y_{\Delta E}y_{E}}{g_{E}\zeta_{E}}\right)-2\left(y_{E}^{\dagger}y_{E}\right)\frac{Y_{\Delta H}}{g_{H}\zeta_{H}}\right],
\end{eqnarray}
%%%
where we have made used of eq. \eqref{eq:expansion1}.
Similarly, we obtain the evolution equation of $\bar{\ell}$ by changing the sign of all chemical potentials
%%%
\begin{eqnarray}
s\frac{dY_{\bar{\ell}}}{dt} & \sim & \frac{1}{2Y^{{\rm nor}}}\left[\left\{ y_{E}^{\dagger}y_{E},\frac{Y_{\Delta\ell}}{g_{\ell}\zeta_{\ell}}\right\}-2\left(\frac{y_{E}^{\dagger}Y_{\Delta E}y_{E}}{g_{E}\zeta_{E}}\right)-2\left(y_{E}^{\dagger}y_{E}\right)\frac{Y_{\Delta H}}{g_{H}\zeta_{H}}\right].
\end{eqnarray}
%%%
Hence the evolution equation for $Y_{\Delta\ell}=Y_{\ell}-Y_{\bar{\ell}}$
has the following form
%%%
\begin{eqnarray}
s\frac{dY_{\Delta\ell}}{dt} & \sim & -\frac{1}{Y^{{\rm nor}}}\left[\left\{ y_{E}^{\dagger}y_{E},\frac{Y_{\Delta\ell}}{g_{\ell}\zeta_{\ell}}\right\} -2\frac{y_{E}^{\dagger}Y_{\Delta E}y_{E}}{g_{E}\zeta_{E}}-2y_{E}^{\dagger}y_{E}\frac{Y_{\Delta H}}{g_{H}\zeta_{H}}\right].
\end{eqnarray}
%%%
Repeating the exercise above, we obtain the evolution equation for
$Y_{\Delta E} = Y_E - Y_{\bar E}$ with the following flavor structure:
%%%
\begin{eqnarray}
s\frac{dY_{\Delta E}}{dt} & \sim & -\frac{1}{Y^{{\rm nor}}}\left[\left\{ y_{E}y_{E}^{\dagger},\frac{Y_{\Delta E}}{g_{E}\zeta_{E}}\right\} -2\frac{y_{E}Y_{\Delta\ell}y_{E}^{\dagger}}{g_{\ell}\zeta_{\ell}}+2y_{E}y_{E}^{\dagger}\frac{Y_{\Delta H}}{g_{H}\zeta_{H}}\right].
\end{eqnarray}
%%%
Under rotations in flavor spaces $E \to UE$, $\ell \to V\ell$, $y_E \to U y_E V^\dagger$, the kinetic equations above will have the same form (flavor covariant) since $Y_{\Delta \ell} \to V Y_{\Delta \ell} V^\dagger$ and $Y_{\Delta E} \to U Y_{\Delta E} U^\dagger$.

As a final remark, in a radiation-dominated Universe and assuming entropy conservation, we can trade the time variable with
temperature $T$ using the relation 
%%%
\begin{eqnarray}
\frac{dz}{dt} & = & z{\cal H},
\end{eqnarray}
%%%
where we have defined $z\equiv\frac{M_{{\rm ref}}}{T}$ with $M_{{\rm ref}}$
an arbitrary mass scale.

\section{Quark number asymmetries \label{app:quark_relations}}

Here, we list the relations between quark number asymmetries with $Y_{\widetilde{\Delta}}$
and $Y_{\Delta E}$ assuming all other conserved charges are zero.
We have included the contributions from possible new scalar fields $\phi_{i}$
with hypercharge $q_{\phi_{i}}^{Y}$. For $T>T_{u}$, all quark number
asymmetries are independent of $Y_{\widetilde{\Delta}}$, $Y_{\Delta E}$
and $Y_{\Delta\phi_{i}}$.

For $T_{u}<T<T_{t}$, we have
%%%
\begin{eqnarray}
Y_{\Delta Q_{3}} & = & \frac{1}{3}\left({\rm Tr}Y_{\widetilde{\Delta}}-2{\rm Tr}Y_{\Delta E}+2\sum_{i}q_{\phi_{i}}^{Y}Y_{\Delta\phi_{i}}\right),\\
Y_{\Delta t} & = & -\frac{1}{3}\left({\rm Tr}Y_{\widetilde{\Delta}}-2{\rm Tr}Y_{\Delta E}+2\sum_{i}q_{\phi_{i}}^{Y}Y_{\Delta\phi_{i}}\right),
\end{eqnarray}
%%%
while the rest are independent of $Y_{\widetilde \Delta}$, $Y_{\Delta E}$ and
$Y_{\Delta\phi_{i}}$.

For $T_{B}<T<T_{u}$, we have
%%%
\begin{eqnarray}
Y_{\Delta Q_{1}} & = & Y_{\Delta Q_{2}}=-\frac{3}{23}\left({\rm Tr}Y_{\widetilde{\Delta}}-2{\rm Tr}Y_{\Delta E}+2\sum_{i}q_{\phi_{i}}^{Y}Y_{\Delta\phi_{i}}\right),\\
Y_{\Delta Q_{3}} & = & \frac{6}{23}\left({\rm Tr}Y_{\widetilde{\Delta}}-2{\rm Tr}Y_{\Delta E}+2\sum_{i}q_{\phi_{i}}^{Y}Y_{\Delta\phi_{i}}\right),\\
Y_{\Delta u} & = & Y_{\Delta d}=Y_{\Delta c}=Y_{\Delta s}=Y_{\Delta b}=\frac{3}{46}\left({\rm Tr}Y_{\widetilde{\Delta}}-2{\rm Tr}Y_{\Delta E}+2\sum_{i}q_{\phi_{i}}^{Y}Y_{\Delta\phi_{i}}\right),\\
Y_{\Delta t} & = & -\frac{15}{46}\left({\rm Tr}Y_{\widetilde{\Delta}}-2{\rm Tr}Y_{\Delta E}+2\sum_{i}q_{\phi_{i}}^{Y}Y_{\Delta\phi_{i}}\right).
\end{eqnarray}
%%%

For $T_{u-b}<T<T_{B}$, we have
%%%
\begin{eqnarray}
Y_{\Delta Q_{1}} & = & Y_{\Delta Q_{2}}=\frac{2}{115}\left(4{\rm Tr}Y_{\widetilde{\Delta}}+15{\rm Tr}Y_{\Delta E}-15\sum_{i}q_{\phi_{i}}^{Y}Y_{\Delta\phi_{i}}\right),\\
Y_{\Delta Q_{3}} & = & \frac{1}{115}\left(53{\rm Tr}Y_{\widetilde{\Delta}}-60{\rm Tr}Y_{\Delta E}+60\sum_{i}q_{\phi_{i}}^{Y}Y_{\Delta\phi_{i}}\right),\\
Y_{\Delta u} & = & Y_{\Delta d}=Y_{\Delta c}=Y_{\Delta s}=Y_{\Delta b}=\frac{1}{115}\left(19{\rm Tr}Y_{\widetilde{\Delta}}-15{\rm Tr}Y_{\Delta E}+15\sum_{i}q_{\phi_{i}}^{Y}Y_{\Delta\phi_{i}}\right),\\
Y_{\Delta t} & = & -\frac{1}{115}\left(26{\rm Tr}Y_{\widetilde{\Delta}}-75{\rm Tr}Y_{\Delta E}+75\sum_{i}q_{\phi_{i}}^{Y}Y_{\Delta\phi_{i}}\right).
\end{eqnarray}
%%%

For $T_{u-c}<T<T_{u-b}$, we have
%%%
\begin{eqnarray}
Y_{\Delta Q_{1}} & = & Y_{\Delta Q_{2}}=Y_{\Delta Q_{3}}=\frac{1}{5}{\rm Tr}Y_{\widetilde{\Delta}},\\
Y_{\Delta u} & = & Y_{\Delta d}=Y_{\Delta c}=Y_{\Delta s}=\frac{1}{10}{\rm Tr}Y_{\widetilde{\Delta}},\\
Y_{\Delta t} & = & -\frac{1}{5}\left({\rm Tr}Y_{\widetilde{\Delta}}-3{\rm Tr}Y_{\Delta E}+3\sum_{i}q_{\phi_{i}}^{Y}Y_{\Delta\phi_{i}}\right),\\
Y_{\Delta b} & = & \frac{1}{5}\left(2{\rm Tr}Y_{\widetilde{\Delta}}-3{\rm Tr}Y_{\Delta E}+3\sum_{i}q_{\phi_{i}}^{Y}Y_{\Delta\phi_{i}}\right).
\end{eqnarray}
%%%

For $T_{B_{3}-B_{2}}<T<T_{u-c}$, we have
%%%
\begin{eqnarray}
Y_{\Delta Q_{1}} & = & \frac{1}{130}\left(11{\rm Tr}Y_{\widetilde{\Delta}}+30{\rm Tr}Y_{\Delta E}-30\sum_{i}q_{\phi_{i}}^{Y}Y_{\Delta\phi_{i}}\right),\\
Y_{\Delta Q_{2}} & = & \frac{1}{130}\left(41{\rm Tr}Y_{\widetilde{\Delta}}-30{\rm Tr}Y_{\Delta E}+30\sum_{i}q_{\phi_{i}}^{Y}Y_{\Delta\phi_{i}}\right),\\
Y_{\Delta Q_{3}} & = & \frac{1}{5}{\rm Tr}Y_{\widetilde{\Delta}},\\
Y_{\Delta u} & = & Y_{\Delta d}=Y_{\Delta s}=\frac{1}{260}\left(41{\rm Tr}Y_{\widetilde{\Delta}}-30{\rm Tr}Y_{\Delta E}+30\sum_{i}q_{\phi_{i}}^{Y}Y_{\Delta\phi_{i}}\right),\\
Y_{\Delta c} & = & -\frac{1}{260}\left(19{\rm Tr}Y_{\widetilde{\Delta}}-90{\rm Tr}Y_{\Delta E}+90\sum_{i}q_{\phi_{i}}^{Y}Y_{\Delta\phi_{i}}\right),\\
Y_{\Delta t} & = & -\frac{1}{130}\left(17{\rm Tr}Y_{\widetilde{\Delta}}-60{\rm Tr}Y_{\Delta E}+60\sum_{i}q_{\phi_{i}}^{Y}Y_{\Delta\phi_{i}}\right),\\
Y_{\Delta b} & = & \frac{1}{130}\left(43{\rm Tr}Y_{\widetilde{\Delta}}-60{\rm Tr}Y_{\Delta E}+60\sum_{i}q_{\phi_{i}}^{Y}Y_{\Delta\phi_{i}}\right).
\end{eqnarray}
%%%

For $T_{u-s}<T<T_{B_{3}-B_{2}}$, we have
%%%
\begin{eqnarray}
Y_{\Delta Q_{1}} & = & \frac{1}{10}\left({\rm Tr}Y_{\widetilde{\Delta}}+2{\rm Tr}Y_{\Delta E}-2\sum_{i}q_{\phi_{i}}^{Y}Y_{\Delta\phi_{i}}\right),\\
Y_{\Delta Q_{2}} & = & Y_{\Delta Q_{3}}=\frac{1}{20}\left(5{\rm Tr}Y_{\widetilde{\Delta}}-2{\rm Tr}Y_{\Delta E}+2\sum_{i}q_{\phi_{i}}^{Y}Y_{\Delta\phi_{i}}\right),\\
Y_{\Delta u} & = & Y_{\Delta d}=Y_{\Delta s}=\frac{1}{20}\left(3{\rm Tr}Y_{\widetilde{\Delta}}-2{\rm Tr}Y_{\Delta E}+2\sum_{i}q_{\phi_{i}}^{Y}Y_{\Delta\phi_{i}}\right),\\
Y_{\Delta c} & = & Y_{\Delta t}=-\frac{1}{10}\left({\rm Tr}Y_{\widetilde{\Delta}}-4{\rm Tr}Y_{\Delta E}+4\sum_{i}q_{\phi_{i}}^{Y}Y_{\Delta\phi_{i}}\right),\\
Y_{\Delta b} & = & \frac{1}{20}\left(7{\rm Tr}Y_{\widetilde{\Delta}}-10{\rm Tr}Y_{\Delta E}+10\sum_{i}q_{\phi_{i}}^{Y}Y_{\Delta\phi_{i}}\right).
\end{eqnarray}
%%%

For $T_{u-d}<T<T_{u-s}$, we have
%%%
\begin{eqnarray}
Y_{\Delta Q_{1}} & = & Y_{\Delta Q_{2}}=Y_{\Delta Q_{3}}=\frac{1}{5}{\rm Tr}Y_{\widetilde{\Delta}},\\
Y_{\Delta u} & = & Y_{\Delta d}=\frac{1}{10}{\rm Tr}Y_{\widetilde{\Delta}},\\
Y_{\Delta c} & = & Y_{\Delta t}=-\frac{1}{80}\left(7{\rm Tr}Y_{\widetilde{\Delta}}-30{\rm Tr}Y_{\Delta E}+30\sum_{i}q_{\phi_{i}}^{Y}Y_{\Delta\phi_{i}}\right),\\
Y_{\Delta s} & = & Y_{\Delta b}=\frac{1}{80}\left(23{\rm Tr}Y_{\widetilde{\Delta}}-30{\rm Tr}Y_{\Delta E}+30\sum_{i}q_{\phi_{i}}^{Y}Y_{\Delta\phi_{i}}\right).
\end{eqnarray}
%%%

Finally, for $T<T_{u-d}$, we have
%%%
\begin{eqnarray}
Y_{\Delta Q_{1}} & = & Y_{\Delta Q_{2}}=Y_{\Delta Q_{3}}=\frac{1}{5}{\rm Tr}Y_{\widetilde{\Delta}},\\
Y_{\Delta u} & = & Y_{\Delta c}=Y_{\Delta t}=-\frac{1}{55}\left(2{\rm Tr}Y_{\widetilde{\Delta}}-15{\rm Tr}Y_{\Delta E}+15\sum_{i}q_{\phi_{i}}^{Y}Y_{\Delta\phi_{i}}\right),\\
Y_{\Delta d} & = & Y_{\Delta s}=Y_{\Delta b}=\frac{1}{55}\left(13{\rm Tr}Y_{\widetilde{\Delta}}-15{\rm Tr}Y_{\Delta E}+15\sum_{i}q_{\phi_{i}}^{Y}Y_{\Delta\phi_{i}}\right).
\end{eqnarray}
%%%

\section{Transition temperatures\label{app:trans_temps}}

To estimate the transition temperature due to the EW sphaleron interaction $T_{B}$,
we define
%%%
\begin{eqnarray}
c_{B}\left(T\right) & = & \frac{3}{2}\frac{Y_{\Delta B}\left(T\right)}{{\rm Tr}Y_{\widetilde{\Delta}}\left(T\right)},
\end{eqnarray}
%%%
where $Y_{\widetilde{\Delta}}=\frac{Y_{\Delta B}}{3}I_{3\times3}-Y_{\Delta\ell}$.
Then, we solve
%%%
\begin{eqnarray}
s{\cal H}z\frac{dY_{\Delta B}}{dz} & = & -\frac{3\gamma_{{\rm EW}}}{4Y^{{\rm nor}}}\left(\frac{{\rm Tr}Y_{\Delta\ell}}{g_{\ell}\zeta_{\ell}}+3\frac{{\rm Tr}Y_{\Delta Q}}{g_{Q}\zeta_{Q}}\right)\nonumber \\
 & = & -\frac{9\gamma_{{\rm EW}}}{16Y^{{\rm nor}}}\left(3Y_{\Delta B}-2{\rm Tr}Y_{\widetilde{\Delta}}\right),
\end{eqnarray}
%%%
where $z\equiv M_{{\rm ref}}/T$ together with eqs. (\ref{eq:BE_YE})
and (\ref{eq:BE_YtildeDelta}) with the condition
%%%
\begin{eqnarray}
Y_{\Delta H} & = & -\frac{14}{23}\left({\rm Tr}Y_{\widetilde{\Delta}}-2{\rm Tr}Y_{\Delta E}\right),
\end{eqnarray}
%%%
which is valid for $T>T_{u-b}$. For $\alpha_{2}$, the RGE at one loop is \cite{Gross:1973id,Politzer:1973fx}
%%%
\begin{eqnarray}
\alpha_{2}\left(\mu\right) & = & \frac{12\pi\alpha_{2}\left(m_{Z}\right)}{12\pi-19\alpha_{2}\left(m_{Z}\right)+19\alpha_{2}\left(m_{Z}\right)\ln\mu},
\end{eqnarray}
%%%
where we take $\mu=2\pi T$ and $\alpha_{2}\left(m_{Z}\right)=0.0337$
with $m_{Z}=91.2$ GeV. Here we ignore the milder RGE of the charged lepton Yukawa and fix it in the flavor basis
to be $\hat{y}_{E}={\rm diag}\left(2.8\times10^{-6},5.9\times10^{-4},1.0\times10^{-2}\right)$
(the result is independent of basis).

We set $M_{{\rm ref}}=10^{12}$ GeV and for initial conditions, we take $Y_{\Delta B}\left(z_{i}\right)=0$ with $z_{i}=10^{-3}$ and
choose arbitrary values for $Y_{\Delta}\left(z_{i}\right)\sim10^{-10}$
and $Y_{\Delta E}\left(z_{i}\right)\sim10^{-10}$. The behavior of the curve is rather insensitive to a particular choice of $Y_{\Delta}\left(z_{i}\right)$
and $Y_{\Delta E}\left(z_{i}\right)$. Solving up to $z_f = 10^3$, to an accuracy within percent level, one obtains the fitting function
%%%
\begin{eqnarray}
c_{B}\left(T\right) & = & 1-e^{-\frac{T_{B}}{T}},
\end{eqnarray}
%%%
where $T_{B}=2.3\times10^{12}\,{\rm GeV}$.

To estimate the transition temperatures related to quark Yukawa interactions,
one needs to solve for $c_{H}\left(T\right)$ as
%%%
\begin{equation}
c_{H}\left(T\right)=-\frac{Y_{\Delta H}\left(T\right)}{{\rm Tr}Y_{\widetilde{\Delta}}\left(T\right)-2{\rm Tr}Y_{\Delta E}\left(T\right)}.
\end{equation}
%%%
Here $Y_{\Delta H}$ should be treated as an independent variable
and one will need to construct a Boltzmann equation for $Y_{\Delta H}$
taking account all the interactions that change the number of Higgs.

\bibliography{biblio}

\end{document}